\documentclass[journal]{IEEEtran}
\usepackage{amsmath}
\usepackage{cite}
\usepackage{graphicx}
\usepackage{epstopdf}
\usepackage{amsfonts,amsmath,amssymb}
\usepackage{graphicx}
\usepackage{url}
\usepackage{bm}
\usepackage{bbm}
\usepackage{subfigure}
\usepackage{stfloats}
\usepackage{citesort}
\usepackage{color}

\usepackage{algorithmic}
\usepackage{algorithm}

\DeclareMathOperator*{\argmin}{argmin}

\newtheorem{corollary}{\textbf{Corollary}}
\newtheorem{theorem}{\textbf{Theorem}}

\newtheorem{lemma}{\textbf{Lemma}}

\begin{document}

\title{Artificial-Noise-Aided Secure Transmission in Wiretap Channels with Transmitter-Side Correlation}

\author{Shihao~Yan, \IEEEmembership{Member, IEEE,} Xiangyun~Zhou, \IEEEmembership{Member, IEEE,} Nan Yang, \IEEEmembership{Member, IEEE,} Biao He, \IEEEmembership{Member, IEEE,} and Thushara D. Abhayapala, \IEEEmembership{Senior Member, IEEE}

\thanks{S. Yan, X. Zhou, N. Yang, and T. D. Abhayapala are  with the Research School of Engineering, Australia National University, Canberra, ACT, Australia (emails: \{shihao.yan, xiangyun.zhou, nan.yang, thushara.abhayapala\}@anu.edu.au).}
\thanks{B. He is with Department of Electronic and Computer Engineering, Hong Kong University of Science and Technology, Hong Kong (email: eebiaohe@ust.hk).}
\thanks{Part of this work has been accepted by IEEE Globecom 2016 \cite{yan2016correlation}. This work was supported by the Australian Research Council's Discovery Projects (DP150103905).}
}

\markboth{Accepted by IEEE Transactions on Wireless Communications}{Yan \MakeLowercase{\textit{et al.}}: Artificial-Noise-Aided Secure Transmission in Wiretap Channels with Transmitter-Side Correlation}

\maketitle

\begin{abstract}
This work for the first time examines the impact of transmitter-side correlation on the artificial-noise-aided secure transmission, based on which a new power allocation strategy for artificial noise (AN) is devised for physical layer security enhancement. Specifically, we design a correlation-based power allocation (CPA) for AN, of which the optimality in terms of achieving the minimum secrecy outage probability is analytically proved in the large system regime with the number of transmit antennas approaching infinity. In order to fully reveal the benefits of the CPA, we derive easy-to-evaluate expressions for the secrecy outage probability achieved by the CPA.
Our study demonstrates that the CPA is nearly optimal and significantly outperforms the widely-used uniform power allocation (UPA) even for a moderately small number of correlated transmit antennas.
Furthermore, our numerical results reveal a fundamental difference between the CPA and UPA. That is when the number of correlated transmit antennas increases, the secrecy outage probability of the CPA always reduces while the secrecy outage probability of the UPA suffers from a saturation point.
\end{abstract}


\IEEEpeerreviewmaketitle

\section{Introduction}

\subsection{Background and Motivation}

Crucial concerns on the security of wireless communication are emerging due to the fact that a large amount of confidential information (e.g., email/bank account information and password, credit card details) is currently conveyed by the wireless medium. Apart from the traditional cryptographic techniques, physical layer security has recently become another key mechanism for safeguarding wireless communications and thus attracted a high level of research interest due to its two noticeable advantages \cite{bloch2011physical,zhou2013physical,yang2015safeguarding}. The first one lies in the fact that physical layer security can guarantee information secrecy regardless of an eavesdropper's computational capability, which leads to that perfect secrecy can be achieved at the physical layer alone. Second, physical layer security eliminates the centralized key distribution and management requested by cryptographic techniques, thus facilitating the management and improving the efficiency of wireless communications. In pioneering  studies, e.g., \cite{shannon1949communication,wyner1975the}, a
wiretap channel was proposed as the fundamental model of physical layer security, in which an eavesdropper (Eve) wiretaps the wireless communication from a transmitter (Alice) to an intended receiver (Bob).

Motivated by multiple-input multiple-output (MIMO) techniques, physical layer security in MIMO wiretap channels has attracted considerable research interest in the past decade (e.g., \cite{khisti2010secure,mukherjee2011robust,yan2014transmit}). In this context, an increasing amount of research effort has been devoted to the
artificial-noise-aided secure transmission due to its robustness and desirable performance (e.g., \cite{goel2008guaranteeing,zhou2010secure,li2013spatially,zheng2013improving,yang2015artificial,zhang2013on,yang2016optimal,zhang2015artificial}). The utilization of artificial noise (AN) to enhance physical layer security was proposed in \cite{goel2008guaranteeing}, where the AN is isotropically transmitted in the null space of the main channel (i.e., the channel between Alice and Bob) in order to possibly reduce the quality of the eavesdropper's channel (i.e., the channel between Alice and Eve) without causing interference to Bob. In \cite{zhou2010secure}, an achievable secrecy rate was derived in a closed-form expression, based on which the authors optimized the power allocation between the information signal and the AN. It was shown that equal power allocation is a near-optimal strategy in terms of maximizing the secrecy rate. Instead of transmitting AN isotropically,  \cite{li2013spatially} considered joint optimization of the covariance matrices of the information signal and the AN for secrecy rate optimization. It was shown that transmit beamforming is optimal for secrecy rate maximization, which has been proved as optimal without using AN in \cite{khisti2010securep1}. Besides transmitting AN by Alice, a full-duplex Bob can also transmit AN in order to create interference to Eve, and this strategy is highly desirable in some specific practical scenarios (e.g., where Eve is close to Bob but far from Alice) \cite{zheng2013improving}. In \cite{yang2015artificial}, practical transmission schemes (e.g., on-off transmission) with AN were examined and associated system parameters were optimized to maximize the effective secrecy rate.
Meanwhile, massive MIMO is emerging as a key technology enabler for future 5G wireless networks, and thus physical layer security in the context of massive MIMO is attracting increasing research interests. For example, in \cite{zhu2014secure} the secure downlink transmission in a multicell massive MIMO system was examined, in which two AN shaping matrices were considered. The opportunities and challenges of physical layer security in the context of massive MIMO were discussed in \cite{kapetanovic2015physical}, in which active pilot contamination attacks were revealed to be difficult to detect but of high harmfulness. A most recent work \cite{zhu2016linear} investigated different precoding strategies of data and AN in secure massive MIMO systems and the impact of channel estimation, pilot contamination, multicell interference, and path-loss was examined as well.

In many practical scenarios, correlation exists among the multiple antennas at one transceiver due to limited separation between antenna elements or poor scattering conditions. This motivates the intensive investigation of how the antenna correlation affects the capacity of traditional MIMO systems without secrecy constraints  (see \cite{hanlen2012capacity} and reference therein). For example, in \cite{hanlen2012capacity} it was shown that the transmit antenna correlation may increase capacity while the receive antenna correlation always leads to capacity reduction. It was proved in \cite{jafar2005multiple} that the link capacity is independent of the eigenvectors of correlation matrices and the capacity of a MIMO channel with channel state information (CSI) at the receiver increases almost surely as the number of transmit antennas increases. However, in the context of physical layer security the study on correlation is still in its infancy. The impact of receiver-side correlation at Bob and Eve on the transmit antenna selection (TAS) at Alice was investigated in wiretap channels \cite{yang2013physical}. It was shown that the receiver-side correlation enhances secrecy performance of TAS in the low average signal-to-noise ratio (SNR) regime of the main channel, but leads to performance degradation in the medium and high average SNR regimes \cite{yang2013physical}. Furthermore, spatial receiver-side correlation was introduced into wiretap channels in order to study physical layer security without knowing the number of eavesdropper's antennas \cite{he2015achieving}. However, the impact of the \emph{transmitter-side correlation} has never been examined in the context of physical layer security, not to mention the AN-aided secure transmission. This leaves an important gap in our understanding on the performance of the AN-aided secure transmission, and closing this gap forms the core of this work.

\subsection{Our Main Contributions}

In this work, we first detail the secure transmission with AN in wiretap channels with transmitter-side correlation, based on which we determine the optimal power allocation (OPA) for AN that minimizes the secrecy outage probability as an $(N_t-1)$-dimensional numerical search problem (where $N_t$ is the number of antennas at Alice). Then, focusing on the large system regime with $N_t \rightarrow \infty$ we derive a closed-form solution to the optimal power allocation, named the correlation-based power allocation (CPA), in which Alice allocates all the AN power to one specific direction determined by the transmitter-side correlation matrix and the CSI of the main channel. We further analytically prove that the CPA maximizes the average interference power to Eve for any finite number of correlated transmit antennas. In order to fully understand the benefits of our proposed CPA, we derive easy-to-evaluate expressions for the secrecy outage probability achieved by the CPA, based on which we determine the optimal power allocation between the information signal and AN. Based on the conducted analysis, we draw useful insights on the comparison between the proposed CPA and the widely-used uniform power allocation (UPA), in which AN is isotropically transmitted in the null space of the main channel \cite{goel2008guaranteeing,zhou2010secure,li2013spatially,zheng2013improving,yang2015artificial,zhang2013on,yang2016optimal,zhang2015artificial}.

We present numerical results to characterize the secrecy performance of the CPA with the OPA and UPA as the benchmarks. Our results first demonstrate that a moderate level of antenna correlation already allows the CPA to achieve the nearly optimal secrecy performance even with a small number of transmit antennas. In such a situation, the CPA significantly outperforms the UPA. Our results also reveal a fundamental difference between the CPA and UPA. That is when the number of correlated transmit antennas (i.e., $N_t$) increases, the secrecy outage probability achieved by the CPA always decreases, while the secrecy outage probability achieved by the UPA suffers from a saturation point. Furthermore, our numerical studies show that the transmitter-side correlation decreases the performance of the AN-aided secure transmission, which is mainly due to the fact that the null space of the main channel disappears as transmit antennas become fully correlated.

\subsection{Paper Organization and Notation}

The rest of this paper is organized as follows. Section \ref{system_model} details our system model and the AN-aided secure transmission in correlated fading channels. Section~\ref{power_allocation} presents the three power allocations (i.e., OPA, CPA, and UPA), analytically proves the optimality of the CPA in the large system regime, and clarifies the benefits of the CPA relative to the UPA. Section~\ref{performance_analysis} derives the secrecy outage probability achieved by the CPA and optimizes the power allocation between the information signal and AN in the AN-aided secure transmission with CPA. Section~\ref{numerical} provides numerical results to confirm our analysis and provide useful insights on the impact of transmit antenna correlation. Section \ref{conclusion} draws concluding remarks.

\emph{Notations:} Scalar variables are denoted by italic symbols. Vectors and matrices are denoted by lower-case and upper-case boldface symbols, respectively. Given a complex number $z$, $|z|$ denotes the modulus of $z$. Given a complex vector $\mathbf{x}$, $\|\mathbf{x}\|$ denotes the Euclidean norm and $\mathbf{x}^{\dag}$ denotes the conjugate transpose of $\mathbf{x}$. The $L\times L$ identity matrix is referred to as $\mathbf{I}_{L}$ and $\mathbb{E}[\cdot]$ denotes expectation.

\section{System Model}\label{system_model}

\subsection{Channel Model}

\begin{figure}[!t]
    \begin{center}
        \includegraphics[width=0.8\columnwidth]{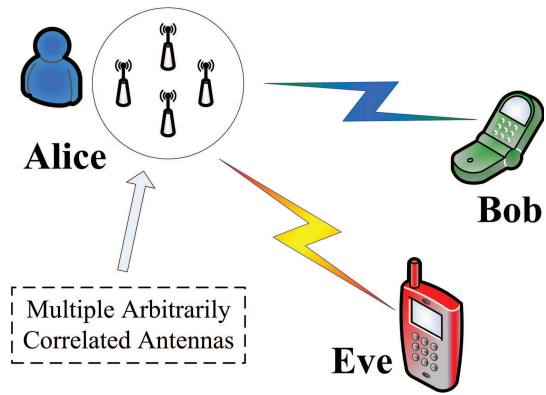}
        \caption{Illustration of the wiretap channel of interest where Alice is equipped with multiple arbitrarily correlated antennas.}
        \label{fig:system}
    \end{center}
\end{figure}

The wiretap channel of interest is illustrated in
Fig.~\ref{fig:system}, where Alice is equipped with $N_t$ correlated antennas, Bob is equipped with a single antenna, and Eve is equipped with a single antenna. This consideration can be justified in the scenarios where Alice is a multi-antenna access point or base station while both Bob and Eve are single-antenna user terminals in the network (and Alice transmits confidential information to Bob which should not be intercepted by Eve).
We assume that all the wireless channels within our system model are subject to quasi-static Rayleigh fading and the average fading power is normalized to one.
Specifically, we adopt the separable correlation model (i.e., Kronecker model) and thus the $1 \times N_t$ main channel (i.e., the channel between Alice and Bob) vector is given by \cite{kermoal2002a}
\begin{align}\label{h_definition}
\mathbf{h} = \mathbf{h}_s \mathbf{T}^{1/2},
\end{align}
where $\mathbf{h}_s \in \mathcal{C}^{1 \times N_t}$ has independent and identically distributed (i.i.d.) circularly-symmetric complex Gaussian entries  (i.e., the entries are i.i.d circularly symmetric complex Gaussian random variables with zero-mean and unit-variance), $\mathbf{T}$ is the transmitter-side correlation matrix, and $\mathbf{T}^{1/2}$ denotes the \emph{cholesky square root} of $\mathbf{T}$. We note that the transmitter-side correlation matrix $\mathbf{T}$ in the Kronecker model is determined by the signal frequency, transmitter-side antenna geometry, and transmitter-side scattering characteristics \cite{kermoal2002a,abdi2002a,byers2004spatially}. As such, in this work we assume that $\mathbf{T}$ is known by Alice and the eavesdropper's channel is subject to the same transmit antenna correlation as the main channel.
Without loss of generality, we assume that $\mathbf{T}$ is a positive symmetric matrix. Thus, its singular value decomposition (SVD) can be written as $\mathbf{T} = \mathbf{U}_T \mathbf{\Lambda}  \mathbf{U}_T^{\dag}$, where $\mathbf{U}_T$ is a unitary matrix and $\mathbf{\Lambda} = \text{diag}[\lambda_1, \cdots, \lambda_{N_t}]$ is the diagonal matrix with $\lambda_i$ as the $i$-th singular value of $\mathbf{T}$. Then, we can rewrite \eqref{h_definition} as $\mathbf{h} = \mathbf{h}_s \sqrt{\mathbf{\Lambda}}\mathbf{U}_T^{\dag}$, where each element of $\sqrt{\mathbf{\Lambda}}$ is the square root of the corresponding element of $\mathbf{\Lambda}$.
We note that in most practical scenarios, Alice will send pilots \emph{a priori} or together with data to enable the estimation of $\mathbf{h}$ at Bob, and then Alice acquires the estimated CSI of the main channel through requesting feedback from Bob. The obtained CSI may suffer from the channel estimation and quantization errors. However, if sufficient resources are available for the channel estimation and feedback (e.g., transmit power and/or time slots are enough for accurate channel estimation), these errors can be made negligible.
As such, we assume that $\mathbf{h}$ is perfectly known at all transceivers in this work, since Eve may eavesdrop on the feedback from Bob to Alice to obtain $\mathbf{h}$.

Since the eavesdropper's channel is also subject to the transmitter-side correlation matrix $\mathbf{T}$, the $1 \times N_t$ eavesdropper's channel vector in the Kronecker model is given by
\begin{align}\label{def_G}
\mathbf{g} = \mathbf{g}_s \mathbf{T}^{1/2} = \mathbf{g}_s \sqrt{\mathbf{\Lambda}}\mathbf{U}_T^{\dag},
\end{align}
where $\mathbf{g}_s \in \mathcal{C}^{1 \times N_t}$ has i.i.d circularly-symmetric complex Gaussian entries. We consider a passive eavesdropping scenario, where Alice does not have the full knowledge of $\mathbf{g}$.

\subsection{Secure Transmission with Artificial Noise}

We next detail the secure transmission with AN in the wiretap channel of interest. In this secure transmission scheme, Alice
transmits an information signal $s_{I}$ in conjunction with
an $(N_t-1)\times1$ AN signal vector
$\mathbf{s}_{N}$ \cite{goel2008guaranteeing,zhou2010secure,li2013spatially,zheng2013improving,yang2015artificial,zhang2013on,yang2016optimal,zhang2015artificial}, where $s_{I}$ and each entry of
$\mathbf{s}_{N}$ have unit variance.
We denote the total transmit power of Alice
by $P_{A}$. The fraction of the power allocated
to $s_{I}$ is $\alpha$ ($0<\alpha\leq 1$), and the remaining power $(1-\alpha)P_A$ is allocated to $\mathbf{s}_N$. In order to transmit $s_{I}$ and $\mathbf{s}_{N}$,
Alice designs an $N_t \times{}N_t$ beamforming matrix $\mathbf{V}$ given
by
\begin{align}\label{beamforming_matrix}
\mathbf{V}=\left[\mathbf{v}_{I}~\mathbf{V}_{N}\right],
\end{align}
where $\mathbf{v}_{I} \in \mathcal{C}^{N_t\times 1}$ is used to transmit $s_{I}$
and $\mathbf{V}_{N} \in \mathcal{C}^{N_t\times (N_t-1)}$ is used to transmit
$\mathbf{s}_{N}$. The aim of $\mathbf{v}_{I}$ is to maximize the instantaneous SNR of the main channel, and thus we have $\mathbf{v}_{I} = \frac{\mathbf{h}^{\dag}}{\|\mathbf{h}\|}$. Meanwhile, $\mathbf{V}_N$ is to degrade the quality of the
eavesdropper's channel by transmitting $\mathbf{s}_{N}$ while perfectly avoiding interference to Bob. As such, $\mathbf{V}_{N}$ consists of $N_t - 1$ orthonomal column vectors in the nullspace of $\textbf{h}^{\dag}$. Then, the $N_t\times1$ transmitted
signal vector at Alice, $\mathbf{x}$, is given by
\begin{align}\label{transmitted_signal}
\mathbf{x}&=\left[\mathbf{v}_{I}~\mathbf{V}_{N}\right]
\left[\begin{array}{cc}\sqrt{\alpha P_A}, &0\\
0, &\sqrt{\frac{(1-\alpha)P_A}{N_t-1}} \sqrt{ \mathbf{\Omega}}\end{array}\right]
\left[\begin{array}{c}
s_{I}\\
\mathbf{s}_{N}\end{array}\right]\notag\\
&=\sqrt{\alpha P_A} \mathbf{v}_{I}s_{I}+\sqrt{\frac{(1-\alpha)P_A}{N_t-1}} \mathbf{V}_{N}\sqrt{\mathbf{\Omega}}\mathbf{s}_{N},
\end{align}
where $\mathbf{\Omega} \in \mathcal{C}^{(N_t-1)\times (N_t-1)}$ is the power allocation matrix for $\mathbf{s}_N$ satisfying $\text{tr}(\mathbf{\Omega}) = N_t-1$.

Following \eqref{transmitted_signal} and noting $\mathbf{h}\mathbf{V}_{N}= 0$, the received signal at Bob
is given by
\begin{align}\label{Receive_Bob}
y=\mathbf{h}\mathbf{x}+n_{B}=\sqrt{\alpha P_A} \mathbf{h}\mathbf{v}_{I}s_{I}+n_{B},
\end{align}
where $n_{B}$ is the additive white Gaussian noise (AWGN) at
Bob satisfying
$\mathbb{E}[n_{B}n_{B}^{\dag}]=\sigma_{B}^{2}$.
Based on \eqref{Receive_Bob}, the instantaneous SNR at Bob
is given by
\begin{align}\label{SNR_Bob}
\gamma_{B}=\alpha \mu_B\left\|\mathbf{h}\right\|^{2} = \alpha \mu_B \sum_{i = 1}^{N_t} \lambda_i |\mathbf{h}_s(i)|^2,
\end{align}
where $\mu_B=P_{A}/\sigma_{B}^{2}$. We note that $\gamma_B$ is a function of $\mathbf{\Lambda}$ but not a function of $\mathbf{U}_T$.


Following \eqref{transmitted_signal}, the received signal at Eve
is given by
\begin{align}\label{Receive_Eve}
z&=\mathbf{g}\mathbf{x}+n_{E}\notag\\
&=\sqrt{\alpha P_A} \mathbf{g}\mathbf{v}_{I}s_{I}+
\sqrt{\frac{(1-\alpha)P_A}{N_t-1}} \mathbf{g} \mathbf{V}_{N}\sqrt{\mathbf{\Omega}} \mathbf{s}_N+{n}_{E},
\end{align}
where ${n}_{E}$ is the AWGN at
Eve satisfying
$\mathbb{E}[{n}_{E}{n}_{E}^{\dag}]
=\sigma_{E}^{2}$. It is crucial to clarify
that although Eve knows $\mathbf{h}$ and $\mathbf{V}$, she cannot eliminate the interference
caused by $\mathbf{V}_{N}\mathbf{s}_{N}$ due to $N_t>1$.
Following \eqref{Receive_Eve}, the instantaneous SINR at Eve is given by \cite{zhou2010secure}
\begin{align}\label{SNR_Eve}
\gamma_{E}=\alpha \mathbf{g}\mathbf{v}_{I} \left(\frac{1-\alpha}{N_t-1}\mathbf{g}
\mathbf{V}_{N}\mathbf{\Omega}\mathbf{V}_{N}^{\dag}\mathbf{g}^{\dag}+\frac{1}
{\mu_E}\right)^{-1}\mathbf{v}_{I}^{\dag}\mathbf{g}^{\dag},
\end{align}
where $\mu_E=P_{A}/\sigma_{E}^{2}$. We note that $\gamma_E$ is a function of $\mathbf{\Lambda}$ but not of $\mathbf{U}_T$ (the proof is similar to that of Lemma 2 in \cite{hanlen2012capacity} and omitted here). Noting $\gamma_B$  is a function of $\mathbf{\Lambda}$ but not of $\mathbf{U}_T$ as well, we can conclude that only the singular values of the correlation matrix $\mathbf{T}$ have impact on the secure communication with AN, while the eigenvectors of $\mathbf{T}$ (i.e., $\mathbf{U}_T$) have no impact.

\subsection{Secrecy Performance Metric}

Since the CSI of the main channel is assumed to be known by Alice, the capacity of the main channel, $C_B = \log_2(1+\gamma_B)$, is available at Alice. On the other hand, Alice does not know the capacity of the eavesdropper's channel,  $C_E = \log_2(1+\gamma_E)$, due to the fact that she cannot access the CSI of the eavesdropper's channel. Considering quasi-static fading channels, we adopt the secrecy outage probability as our key performance metric, which is defined as the probability that the target rate of a secure transmission is larger than the secrecy capacity. The secrecy outage probability is given by \cite{bloch2008wireless}
\begin{align}\label{Pso_definition}
P(R_s) &= \Pr(C_s \!<\! R_s) = \Pr(C_B - R_s \!<\! C_E),
\end{align}
where $R_s$ is the target rate of a secure transmission and $C_s = [C_B - C_E]^{+}$ is the secrecy capacity, where $[x]^{+} = \max\{0,x\}$.
In order to facilitate the secure transmission design under a given main channel condition, we study the secrecy outage probability in \eqref{Pso_definition} for a given $C_B$, and this outage is solely caused by the uncertainty of $C_E$. We also note that if $R_s \geq C_B$ the main channel cannot support such a secure transmission (i.e., the secrecy outage probability is one).

\section{Power Allocation for Artificial Noise}\label{power_allocation}

In this section, we first present the OPA (i.e., optimal power allocation) for AN in the secure transmission that minimizes the secrecy outage probability in the wiretap channel with transmitter-side correlation. Then, focusing on the large system regime with $N_t \rightarrow \infty$, we derive a closed-form solution to the optimal power allocation based on the correlation matrix, named CPA (i.e., correlation-based power allocation). In addition, we discuss the UPA (i.e., uniform power allocation) in the wiretap channel with transmitter-side correlation as a benchmark in this section.

\subsection{Optimal Power Allocation}

Utilizing the secrecy outage probability as the objective function, the optimization problem of power allocation for AN can be written as
\begin{equation}\label{original_optimization}
\begin{split}
\min_{\mathbf{\Omega}}~~P(R_s),~~
\text{s.t.} ~~ \text{tr}(\mathbf{\Omega}) = N_t-1.
\end{split}
\end{equation}
The optimization problem presented in \eqref{original_optimization} involves the determination of $(N_t-1)^2$ complex entries of the  power allocation matrix $\mathbf{\Omega}$ (i.e., $2(N_t-1)^2$ real numbers), which is of high complexity. We have the following lemma to simplify \eqref{original_optimization} as a $(N_t-1)$-dimensional numerical search problem. For the sake of clear presentation, we define a positive definite Hermitian matrix as
\begin{align}
\mathbf{Q} = \mathbf{V}_{N}^{\dag} \mathbf{T} \mathbf{V}_{N},
\end{align}
and its SVD can be written as
\begin{align}
\mathbf{Q} = \mathbf{W}\Theta \mathbf{W}^{\dag},
\end{align}
where $\mathbf{W}$ is a unitary matrix and $\Theta = \text{diag}[\theta_1, \dots, \theta_{N_t - 1}]$ is the diagonal matrix with $\theta_m$ as the $m$-th singular value of $\mathbf{Q}$ with $\theta_1 \geq \theta_2 \geq \dots \geq \theta_{N_t-1}$. We now present Lemma~\ref{lemma1} based on the SVD of $\mathbf{Q}$.

\begin{lemma}\label{lemma1}
The optimization problem presented in \eqref{original_optimization} can be simplified as
\begin{equation}\label{simplified_optimization}
\begin{split}
\min_{\mathbf{\Phi}}~~&P(R_s),\\
\text{s.t.} ~~& \mathbf{\Omega} = \mathbf{W}\mathbf{\Phi}\mathbf{W}^{\dag}, ~~\text{tr}(\mathbf{\Phi}) = N_t-1,\\
& \mathbf{\Phi} = \text{diag}\left[\phi_1, \phi_2, \dots, \phi_{N_t-1}\right],
\end{split}
\end{equation}
where $\phi_1, \phi_2, \dots, \phi_{N_t-1}$ are non-negative real numbers.
\end{lemma}
\begin{IEEEproof}
We note that the selection of $\mathbf{\Omega}$
affects $P(R_s)$ only through $\mathbf{g}
\mathbf{V}_{N}\mathbf{\Omega}\mathbf{V}_{N}^{\dag}\mathbf{g}^{\dag}$ involved in $\gamma_E$ given in \eqref{SNR_Eve}.  Based on the definition of $\mathbf{g}$ given in \eqref{def_G}, we have $\mathbf{V}_{N}^{\dag}\mathbf{g}^{\dag} \sim \mathcal{CN}(\mathbf{0}, \mathbf{Q})$. Then we know that $\mathbf{g}\mathbf{V}_{N}\mathbf{W}$ has i.i.d circularly-symmetric complex Gaussian entries. As proved in \cite[Theorem~1 and Lemma~1]{jafar2004transmitter}, $\mathbf{g}\mathbf{V}_{N}\mathbf{W}\mathbf{\Phi}\mathbf{W}^{\dag}\mathbf{V}_{N}^{\dag}\mathbf{g}^{\dag}$ is equal in distribution to $\mathbf{g}
\mathbf{V}_{N}\mathbf{\Omega}\mathbf{V}_{N}^{\dag}\mathbf{g}^{\dag}$ for general $\mathbf{\Phi}$ and $\mathbf{\Omega}$. This completes the proof of Lemma~\ref{lemma1}.
\end{IEEEproof}

We note that the optimization problem presented in \eqref{simplified_optimization} is much less complex than that provided in \eqref{original_optimization}. This is due to the fact that in \eqref{original_optimization} there are $2(N_t -1)^2$ real numbers to determine for $\mathbf{\Omega}$ while we only have to determine $N_t-1$ real numbers for $\mathbf{\Phi}$ in \eqref{simplified_optimization}.
We also note that analytical solution to \eqref{simplified_optimization} is still mathematically intractable. This is mainly due to the fact that $P(R_s)$ cannot be derived in a closed-form expression for a general $\mathbf{\Phi}$. The difficulty lies in the fact that $\mathbf{g}\mathbf{V}_{N}\mathbf{W}\mathbf{\Phi}\mathbf{W}^{\dag}\mathbf{V}_{N}^{\dag}\mathbf{g}^{\dag}$ and $|\mathbf{g}\mathbf{v}_I|^2$ are correlated in the expression of $\gamma_E$ given in  \eqref{SNR_Eve}, which leads to that the probability density function (pdf) of $\gamma_E$ is mathematically intractable. Therefore, the optimization problem given in \eqref{simplified_optimization} can only be solved through numerical simulations, which is of high complexity and time-consuming for large $N_t$. As such, in the following we develop a sub-optimal but much simpler power allocation, and analytically prove its optimality in the large system regime with $N_t \rightarrow \infty$.

\subsection{Correlation-Based Power Allocation}

Now, we propose the CPA, which is optimal in terms of minimizing $P(R_s) $ in the large system regime with $N_t \rightarrow \infty$, in the following theorem.
\begin{theorem}\label{theorem1}
As $N_t \rightarrow \infty$, the optimal solution to $\mathbf{\Omega}$ that minimizes $P(R_s)$ is given by
\begin{align}\label{spatial_optimization}
\mathbf{\Omega}^{\ast} = (N_t-1)\mathbf{w}_I\mathbf{w}_I^{\dag},
\end{align}
where $\mathbf{w}_I$ is the principal eigenvector corresponding
to the largest singular value of $\mathbf{Q}$ (i.e., $\mathbf{\Omega}^{\ast} = \mathbf{W}\mathbf{\Phi}^{\ast}\mathbf{W}^{\dag}$ and $\mathbf{\Phi}^{\ast} = \text{diag}[N_t-1, 0, \dots, 0]$).
\end{theorem}
\begin{IEEEproof}
Due to the distance concentration phenomenon \cite{francois2007the}, when $N_t \rightarrow \infty$ both $|\mathbf{g}\mathbf{v}_{I}|^2$ and $\|\mathbf{g}\mathbf{V}_{N}\mathbf{W}\sqrt{\mathbf{\Phi}}\|^2$ involved in $\gamma_E$ approach their mean values.
We note that the mean of $\|\mathbf{g}\mathbf{V}_{N}\mathbf{W}\sqrt{\mathbf{\Phi}}\|^2$ is positive and $\mu_E \geq 0$. As such, we can conclude that $\gamma_E$ approaches its mean as $N_t \rightarrow \infty$ based on the quotient rule for limits, which states that the limit of a quotient of two functions is the quotient of their limits, provided the limit of the denominator is not zero \cite{voiculescu1991limit}.
It follows that the minimization of the secrecy outage probability $P(R_s)$ is equivalent to minimizing the mean of $\gamma_E$. We note that $\mathbf{\Phi}$ only varies the value of $\|\mathbf{g}\mathbf{V}_{N}\mathbf{W}\sqrt{\mathbf{\Phi}}\|^2$ (i.e., $|\mathbf{g}\mathbf{v}_{I}|^2$ is not a function of $\mathbf{\Phi}$). Therefore, $\mathbf{\Phi}$ is to maximize the mean of $\|\mathbf{g}\mathbf{V}_{N}\mathbf{W}\sqrt{\mathbf{\Phi}}\|^2$ in order to minimize $P(R_s)$ as per the expression of $\gamma_E$ given in \eqref{SNR_Eve}. As mentioned in the proof of Lemma~\ref{lemma1}, $\mathbf{g}\mathbf{V}_{N}\mathbf{W}$ has i.i.d entries, and thus we have
\begin{align}\label{iid_spa}
\|\mathbf{g}\mathbf{V}_{N}\mathbf{W}\sqrt{\mathbf{\Phi}}\|^2  = \sum_{m = 1}^{N_t - 1} \phi_m \theta_m |\mathbf{g}_I(m)|^2,
\end{align}
where $\mathbf{g}_I = \mathbf{g}\mathbf{V}_{N}\mathbf{W}(\sqrt{\Theta})^{-1}$ has i.i.d circularly-symmetric complex Gaussian entries with unit variance. Then, the mean of $\|\mathbf{g}\mathbf{V}_{N}\mathbf{W}\sqrt{\mathbf{\Phi}}\|^2$ is given by
\begin{align}\label{mean_gvwo}
\mathbb{E}\left[\|\mathbf{g}\mathbf{V}_{N}\mathbf{W}\sqrt{\mathbf{\Phi}}\|^2\right]= \sum_{m = 1}^{N_t - 1} \phi_m \theta_m.
\end{align}
Noting $\theta_1 \geq \theta_2 \geq \dots \geq \theta_{N_t-1}$, in order to maximize $\mathbb{E}[\|\mathbf{g}\mathbf{V}_{N}\mathbf{W}\sqrt{\mathbf{\Omega}}\|^2]$ subject to $\text{tr}(\mathbf{\Phi}) = N_t-1$ (i.e., $\phi_1 + \phi_2 + \dots + \phi_{N_t-1} = N_t-1$), we have to set $\phi_1^{\ast} = N_t-1$ and $\phi_k^{\ast} =0$ for $k = 2, 3, \dots, N_t-1$ (i.e., $\mathbf{\Phi}^{\ast} = \text{diag}[N_t-1, 0,\dots, 0]$).
We note that for $\mathbf{\Phi}^{\ast} = \text{diag}[N_t-1, 0,\dots, 0]$ we have $\mathbf{\Omega}^{\ast} = \mathbf{W}\mathbf{\Phi}^{\ast}\mathbf{W}^{\dag}  = (N_t\!-\!1)\mathbf{w}_I\mathbf{w}_I^{\dag}$. This completes the proof of Theorem~\ref{theorem1}.
\end{IEEEproof}

We note that the CPA allocates all the AN power to the direction corresponding to the largest singular value of $\mathbf{Q}$, which is similar to the beamforming strategy based on $\mathbf{Q}$. The intuitive meaning of the CPA is that Alice first maps the $N_t$-dimensional eavesdropper's channel vector into the $(N_t-1)$-dimensional nullspace of the main channel by applying $\mathbf{V}_N$ and then transmits AN along the average strongest direction of the effective eavesdropper's channel vector $\mathbf{g}\mathbf{V}_N$. This is due to the fact that $\mathbf{Q} = \mathbf{V}_{N}^{\dag} \mathbf{T} \mathbf{V}_{N}$ is the covariance matrix of $\mathbf{g}\mathbf{V}_N$ and thus $\mathbf{w}_I$ corresponds to the average strongest direction of $\mathbf{g}\mathbf{V}_N$.
We also note that this is the first work studying the impact of transmitter-side correlation on the design of secure transmission. As such, we focus more on the transmitter side and considered a simple baseline model for receivers, in which Bob and Eve are both single-antenna devices. Moreover, we clarify that our Theorem~\ref{theorem1} is still valid when Eve is equipped with multiple (i.e., $N_e$) uncorrelated antennas. 


The following corollary states another important property of the CPA.
\begin{corollary}
The CPA (i.e., $\mathbf{\Omega}^{\ast} = (N_t\!-\!1)\mathbf{w}_I$) achieves the maximum average interference to Eve for all values of $N_t$, which is given by
\begin{align}\label{interference_m_spa}
\mathbb{E}\left[|\mathbf{g}
\mathbf{V}_{N}\sqrt{\mathbf{\Omega}^{\ast}}|^2\right] = (N_t - 1)\theta_1.
\end{align}
\end{corollary}
\begin{IEEEproof}
Based on Lemma~1, we know that $\mathbb{E}[\|\mathbf{g}\mathbf{V}_{N}\sqrt{\mathbf{\Omega}}\|^2] = \mathbb{E}[\|\mathbf{g}\mathbf{V}_{N}\mathbf{W}\sqrt{\mathbf{\Phi}}\|^2]$. Then, based on the discussion after \eqref{mean_gvwo} in the proof of Theorem~\ref{theorem1} we can conclude that the CPA maximizes the average interference to Eve (i.e., maximizes $\mathbb{E}[\|\mathbf{g}\mathbf{V}_{N}\sqrt{\mathbf{\Omega}}\|^2]$). Finally, substituting $\mathbf{\Omega}^{\ast} = \mathbf{W}\mathbf{\Phi}^{\ast}\mathbf{W}^{\dag}  = (N_t\!-\!1)\mathbf{w}_I \mathbf{w}_I^{\dag}$ into \eqref{mean_gvwo} we achieve the result given in \eqref{interference_m_spa}.
\end{IEEEproof}

As per Theorem~\ref{theorem1}, the instantaneous SINR at Eve of the CPA for a given $\alpha$ is given by
\begin{align}\label{SNR_Eve_CPA}
\gamma_{E}= \frac{\alpha \mu_E|\mathbf{g}\mathbf{v}_{I}|^2}{(1-\alpha) \mu_E |\mathbf{g}
\mathbf{V}_{N}\mathbf{w}_I|^2 + 1}.
\end{align}



\subsection{Uniform Power Allocation}

In this subsection, we present the UPA as a benchmark to clarify the benefits of our proposed CPA. In the UPA, Alice isotropically allocates the transmit power for the AN among all entries of $\mathbf{s}_N$, i.e., $\mathbf{\Omega} = \mathbf{I}_{N_t-1}$.
Following \eqref{SNR_Eve}, the SINR at Eve of the UPA for a given $\alpha$ is given by \cite{yang2015artificial}
\begin{align}\label{SNR_Eve_UPA}
\gamma_{E}^u = \frac{\alpha \mu_E|\mathbf{g}\mathbf{v}_{I}|^2}{\frac{(1-\alpha)\mu_E}{N_t-1}\|\mathbf{g}
\mathbf{V}_{N}\|^2 + 1}.
\end{align}
Following a similar procedure for obtaining \eqref{iid_spa}, we have
\begin{align}\label{iid_upa}
\|\mathbf{g}\mathbf{V}_{N}\|^2  = \sum_{m = 1}^{N_t - 1} \theta_m |\mathbf{g}_I(m)|^2.
\end{align}
Then, the average interference to Eve achieved by the UPA is given by
\begin{align}\label{interference_m_upa}
\mathbb{E}\left[\|\mathbf{g}
\mathbf{V}_{N}\|^2\right] = \sum_{m = 1}^{N_t - 1} \theta_m.
\end{align}

We note that the UPA is widely adopted in the literature in wiretap channels without correlation. This is due to the fact that Alice cannot access the CSI of the eavesdropper's channel and has no information on $\mathbf{g}$ other than its distribution. With regard to the comparison between the UPA and CPA, we have the following remarks.
\begin{itemize}
\item
The UPA maximizes the average interference to Eve in wiretap channels without correlation (i.e., $\mathbb{E}[\|\mathbf{g}\mathbf{V}_{N}\|^2] \geq \mathbb{E}[\mathbf{g}
\mathbf{V}_{N}\mathbf{\Omega}\mathbf{V}_{N}^{\dag}\mathbf{g}^{\dag}]$ when $\mathbf{T} = \mathbf{I}_{N_t}$).
We note that our proposed CPA achieves the same average interference to Eve as the UPA when $\mathbf{T} = \mathbf{I}_{N_t}$ (i.e., $(N_t-1)\theta_1 = \sum_{m = 1}^{N_t - 1} \theta_m$ when $\mathbf{T} = \mathbf{I}_{N_t}$ due to $\theta_1 = \theta_2 = \dots = \theta_{N_t-1}$ for $\mathbf{T} = \mathbf{I}_{N_t}$).

\item
Comparing \eqref{interference_m_spa} and \eqref{interference_m_upa} we can see that the CPA leads to a larger average interference to Eve than the UPA in wiretap channels with transmitter-side correlation. This is due to $\theta_1 \geq \theta_2 \geq \dots \geq \theta_{N_t-1}$, which leads to $(N_t-1)\theta_1 \geq \sum_{m = 1}^{N_t - 1}\theta_m$.

\item
Following the proof of Theorem~\ref{theorem1}, we know that as $N_t \rightarrow \infty$, these average interferences to Eve of the CPA and UPA as given in \eqref{interference_m_spa} and \eqref{interference_m_upa}, respectively, determine the secrecy outage probabilities of the CPA and UPA, respectively. As such, we can conclude that in the large system regime with $N_t \rightarrow \infty$, the CPA leads to a lower $P(R_s)$  relative to the UPA in wiretap channels with transmitter-side correlation.

\item
The gap between $\theta_1$ and $\frac{1}{N_t-1}\sum_{m = 1}^{N_t - 1}\theta_m$ increases as the correlation becomes more severe. As such, our proposed CPA is more desirable in wiretap channels with high transmitter-side correlation.
\end{itemize}

\section{Secrecy Performance Analysis of the Correlation-based Power Allocation}\label{performance_analysis}

In this section we analyze the secrecy performance of the AN-aided secure transmission with CPA in order to fully understand the benefits of the proposed CPA for AN. As mentioned in Section II-C, we study the secrecy outage probability for a given main channel realization in order to facilitate the channel-dependent transmission design.
Specifically, we derive easy-to-evaluate expressions for the secrecy outage probabilities of the CPA. Based on these expressions,
we determine the optimal power allocation between the information signal and the AN. We note that in practice we may not be able to know the receiver noise level at Eve (i.e., $\sigma_E^2$), and thus we first derive the secrecy outage probability for $\sigma_E^2 = 0$ (referred to as the asymptotic secrecy outage probability) and then derive the secrecy outage probability for arbitrary known values of $\sigma_E^2$ with $\sigma_E^2 > 0$ (referred to as the exact secrecy outage probability). We would like to clarify that the asymptotic secrecy outage probability is not a special case of the exact secrecy outage probability.

\subsection{Asymptotic Secrecy Outage Probability}

For $\sigma_E^2 = 0$ (i.e., $\mu_E \rightarrow \infty$) , following \eqref{SNR_Eve_CPA} the SINR at Eve is given by
\begin{align}\label{asy_SNR_Eve_CPA}
\gamma_E^{\infty}= \frac{\alpha |\mathbf{g}\mathbf{v}_{I}|^2}{(1-\alpha)|\mathbf{g}
\mathbf{V}_{N}\mathbf{w}_I|^2}.
\end{align}
We next derive the asymptotic secrecy outage probability achieved by the CPA for $\sigma_E^2 = 0$  in the following theorem. We note that we have $\alpha < 1$ for $\sigma_E^2 = 0$ in order to guarantee the asymptotic secrecy outage probability to be less than one (i.e., the asymptotic secrecy outage probability equals one for $\alpha = 1$).

\begin{theorem}\label{theorem2}
The asymptotic secrecy outage probability achieved by the CPA is given by
\begin{align}\label{asy_Pso_CPA}
P_{\infty}(R_s) &=1\!\!-\!\!\frac{\sigma_n^2 (1\!-\!\rho)}{\sigma_d^2}\sum_{k=0}^{\infty}\frac{(2k\!+\!1)!\sigma_n^{2k}\rho^k}{(k!)^2\sigma_d^{2k}}\sum_{i=0}^k \frac{\binom{k}{i} \left(\!-\!\frac{\sigma_n^2}{\sigma_d^2}\right)^{k\!-\!i}}{i\!-\!2k\!-\!1}\notag\\
& ~~~~\times \left[\left(\frac{\gamma_B + 1}{2^{R_s}} - 1 + \frac{\sigma_n^2}{\sigma_d^2}\right)^{i\!-\!2k\!-\!1} - \left(\frac{\sigma_n^2}{\sigma_d^2}\right)^{i\!-\!2k\!-\!1}\right],
\end{align}
where
\begin{align}
\rho &= \frac{\alpha (1-\alpha)|\mathbf{v}_I^{\dag} \mathbf{T}\mathbf{V}_N\mathbf{w}_I|^2}{4 \sigma_n^2 \sigma_d^2},\\
\sigma_n &= \sqrt{\frac{\alpha}{2}\mathbf{v}_I^{\dag} \mathbf{T}\mathbf{v}_I},\\
\sigma_d &= \sqrt{\frac{1-\alpha}{2}\mathbf{w}_I^{\dag} \mathbf{V}_N^{\dag} \mathbf{T}\mathbf{V}_N\mathbf{w}_I}.
\end{align}
\end{theorem}
\begin{IEEEproof}
Following \eqref{Pso_definition}, the asymptotic secrecy outage probability of the CPA is given by
\begin{align}\label{asy_Pso_CPA_definition}
P_{\infty}(R_s) = \Pr\left(\gamma_E^{\infty} \!>\! \frac{\gamma_B \!+\! 1}{2^{R_s}} \!-\! 1\right) = 1 \!-\! F_{\gamma_E^{\infty}}\left(\frac{\gamma_B \!+\! 1}{2^{R_s}} \!-\! 1\right),
\end{align}
where $F_{\gamma_E^{\infty}}(\cdot)$ is the cumulative distribution function (cdf) of $\gamma_E^{\infty}$. In order to derive the expression of $P_{\infty}(R_s)$, we next derive $F_{\gamma_E^{\infty}}(\cdot)$. We can rewrite $\gamma_E^{\infty}$ as $\gamma_E^{\infty} = \chi_n/\chi_d$, where
\begin{align}
\chi_n &= \alpha |\mathbf{g}\mathbf{v}_{I}|^2, \\
\chi_d &= (1-\alpha)
|\mathbf{g}
\mathbf{V}_{N}\mathbf{w}_I|^2.
\end{align}
The correlation coefficient between $\chi_n$ and $\chi_d$  is $\rho$. We note that $\rho = 0$ if $\mathbf{T} = \mathbf{I}_{N_t}$ (i.e., uncorrelated fading channels) due to the fact that $\mathbf{v}_I$ and $ \mathbf{V}_N$ are orthogonal.
We also note that $\mathbf{v}_I$, $\mathbf{V}_{N}$, and $\mathbf{w}_I$ are known and deterministic since the CSI of the main channel $\mathbf{h}$ and the correlation matrix $\mathbf{T}$ are assumed to be fixed and known. As such, $\mathbf{g}\mathbf{v}_I$ and $\mathbf{g}
\mathbf{V}_N\mathbf{w}_I$ are both complex Gaussian random variables (for given deterministic $\mathbf{v}_I$, $\mathbf{V}_{N}$, and $\mathbf{w}_I$), which leads to that $\chi_n$ and $\chi_d$ are independent when $\rho = 0$. We next focus on the case where $\rho \neq 0$ since we are interested in correlated fading channels in this work. Again, since $\mathbf{g}\mathbf{v}_I$ and $\mathbf{g}
\mathbf{V}_N\mathbf{w}_I$ are both complex Gaussian random variables (for given deterministic $\mathbf{v}_I$, $\mathbf{V}_{N}$, and $\mathbf{w}_I$) , the joint pdf of $\chi_n$ and $\chi_d$ is given by \cite{mallik2003on}
\begin{align}\label{joint_xn_xd}
f_{\chi_n, \chi_d}(x, y) = \frac{e^{-\frac{1}{2(1-\rho)}\left(\frac{x}{\sigma_n^2}+\frac{y}{\sigma_d^2}\right)}}{4 \sigma_n^2 \sigma_d^2 (1-\rho)}
I_0\left(\frac{\sqrt{\rho}}{1-\rho}\frac{\sqrt{xy}}{\sigma_n \sigma_d}\right),
\end{align}
where $I_0(\cdot)$ is the zero-order modified Bessel function of the first kind, which can be expanded as
\begin{align}\label{zero_expansion}
I_0(t) = \sum_{k=0}^{\infty}\frac{t^{2k}}{2^{2k}(k!)^2}.
\end{align}
Then, the pdf of $\gamma_E^{\infty}$ is derived as
\begin{align}\label{pdf_SINR_asy_CPA}
f_{\gamma_E^{\infty}}(z) &= \int_0^{\infty} y f_{\chi_n, \chi_d}(zy, y) dy\notag\\
&\overset{a}{=}\frac{1}{4 \sigma_n^2 \sigma_d^2 (1-\rho)}\sum_{k=0}^{\infty}\frac{\left(\frac{\sqrt{\rho}}{1-\rho}\frac{\sqrt{z}}{\sigma_n \sigma_d}\right)^{2k}}{2^{2k}(k!)^2} \notag \\
& ~~~~\times \int_0^{\infty}y^{2k + 1} e^{-\frac{1}{2(1-\rho)}\left(\frac{z}{\sigma_n^2}+\frac{1}{\sigma_d^2}\right) y} dy\notag\\
&\overset{b}{=} \frac{1}{4 \sigma_n^2 \sigma_d^2 (1-\rho)}\sum_{k=0}^{\infty}\frac{\left(\frac{\sqrt{\rho}}{1-\rho}\frac{\sqrt{z}}{\sigma_n \sigma_d}\right)^{2k}}{2^{2k}(k!)^2} \notag \\
& ~~~~\times (2k+1)! \left[\frac{1}{2(1-\rho)}\left(\frac{z}{\sigma_n^2}+\frac{1}{\sigma_d^2}\right)\right]^{-(2k+2)}\notag\\
& = \frac{\sigma_n^2 (1-\rho)}{\sigma_d^2}\sum_{k=0}^{\infty}\frac{(2k+1)!\sigma_n^{2k}\rho^k}{(k!)^2\sigma_d^{2k}}\frac{z^k}{\left(z + \frac{\sigma_n^2}{\sigma_d^2}\right)^{2k+2}},
\end{align}
where $\overset{a}{=}$ is achieved by substituting \eqref{zero_expansion} into \eqref{joint_xn_xd} and $\overset{b}{=}$ is obtained based on the following identify \cite[Eq. (3.351.3)]{gradshteuin2007table}
\begin{align}
\int_0^{\infty}x^n e^{- \tau x} dx = n!\tau^{-n-1}.
\end{align}
Then, following \eqref{pdf_SINR_asy_CPA} the cdf of $\gamma_E^{\infty}$ can be derived as
\begin{align}\label{cdf_SINR_asy_CPA}
F_{\gamma_E^{\infty}}(z) &= \int_0^z f_{\gamma_E^{\infty}}(x) dx \notag \\
&\overset{c}{=}\frac{\sigma_n^2 (1-\rho)}{\sigma_d^2}\sum_{k=0}^{\infty}\frac{(2k+1)!\sigma_n^{2k}\rho^k}{(k!)^2\sigma_d^{2k}}\notag \\
& ~~~~\times \int_{\frac{\sigma_n^2}{\sigma_d^2}}^{z + \frac{\sigma_n^2}{\sigma_d^2}} y^{-(2k+2)} \left(y-\frac{\sigma_n^2}{\sigma_d^2}\right)^k dy\notag
\end{align}
\begin{align}
&\overset{d}{=} \frac{\sigma_n^2 (1\!-\!\rho)}{\sigma_d^2}\sum_{k=0}^{\infty}\frac{(2k+1)!\sigma_n^{2k}\rho^k}{(k!)^2\sigma_d^{2k}}\sum_{i=0}^k \frac{\binom{k}{i} \left(\!-\!\frac{\sigma_n^2}{\sigma_d^2}\right)^{k-i}}{i-2k-1}\notag\\
& ~~~~\times \left[\left(z + \frac{\sigma_n^2}{\sigma_d^2}\right)^{i\!-\!2k\!-\!1} - \left(\frac{\sigma_n^2}{\sigma_d^2}\right)^{i\!-\!2k\!-\!1}\right],
\end{align}
where $\overset{c}{=}$ is achieved by exchanging variables (i.e., setting $y = x + \sigma_n^2/\sigma_d^2$) and $\overset{d}{=}$ is obtained by applying the following identity  \cite[Eq. (1.111)]{gradshteuin2007table}
\begin{align}\label{bio_expansion}
(x-t)^n = \sum_{i = 0}^n \binom{n}{i}x^i (-t)^{n-i}.
\end{align}
Then, the secrecy outage probability given in \eqref{asy_Pso_CPA} is achieved by substituting \eqref{cdf_SINR_asy_CPA} into \eqref{asy_Pso_CPA_definition}.
\end{IEEEproof}

We note that although the expression of the secrecy outage probability given in  \eqref{asy_Pso_CPA} involves an infinite series (i.e., $\sum_{k = 0}^{\infty}$), it can be approximated by its finite truncation accurately (e.g., utilizing the first $K$ terms, $\sum_{k = 0}^{K}$, as an approximation of $\sum_{k = 0}^{\infty}$). This is due to the fact that \eqref{asy_Pso_CPA} is convergent since the infinite series involved in \eqref{asy_Pso_CPA} arises from the series expansion of the zero-order modified Bessel function of the first kind given in \eqref{zero_expansion}, which is convergent.

\subsection{Exact Secrecy Outage Probability}

In some scenarios, the receiver noise level at Eve (i.e., $\sigma_E^2$) is possible to be known by Alice. For example, Eve could be a normal user served by Alice, but secure communication between Alice and Bob is requested while treating Eve as an eavesdropper. As such, in this subsection we derive the exact secrecy outage probability achieved by the CPA for $\sigma_E^2 >0$, which is presented in the following theorem.

\begin{theorem}\label{theorem3}
The exact secrecy outage probability achieved by the CPA is given by
\begin{align}\label{Pso_CPA_final}
P(R_s)  = \left\{
\begin{array}{ll}
e^{-\frac{\frac{\gamma_B +1}{2^{R_s}}-1}{\mu_E \mathbf{v}_I^{\dag} \mathbf{T}\mathbf{v}_I}}, & \alpha = 1,\\
P_{so}^{\alpha}(R_s), & 0<\alpha < 1,
\end{array}
\right.
\end{align}
where
\begin{align}\label{Pso_CPA}
P_{so}^{\alpha}&(R_s) = 1-\frac{e^{\frac{1}{2(1-\rho)s_d^2}}}{4 s_n^2 s_d^2 (1-\rho)}\sum_{k=0}^{\infty}\frac{\rho^k  (1\!-\!\rho)^{-k}}{2^{k}(k!)^2 s_d^{2k}}
\sum_{i=0}^k \frac{\binom{k}{i}}{ (\!-\!1)^{i\!-\!k}}\notag \\
& \times \sum_{j = 0}^{k+i+1}\frac{ (k+i+1)! }{j! \left[2(1\!-\!\rho)s_n^2\right]^{j\!-\!i\!-\!2}} \sum_{m = 0}^k\binom{k}{m}\left(\!-\!\frac{s_n^2}{s_d^2}\right)^{k\!-\!m}\notag\\
&\times \left[\digamma\left(k+i\!\!-\!\!j\!\!-\!\!m\!+\!1, \frac{1}{2(1\!-\!\rho)s_n^2}, \frac{\gamma_B \!+\! 1}{2^{R_s}} \!-\! 1 + \frac{s_n^2}{s_d^2}\right)\right.\notag \\
&~~~~\left.- \digamma\left(k+i\!\!-\!\!j\!\!-\!\!m\!+\!1, \frac{1}{2(1\!-\!\rho)s_n^2}, \frac{s_n^2}{s_d^2}\right)\right],\\
s_n &= \sqrt{\frac{\alpha  \mu_E}{2}\mathbf{v}_I^{\dag} \mathbf{T}\mathbf{v}_I},\\
s_d &= \sqrt{\frac{(1-\alpha) \mu_E}{2}\mathbf{w}_I^{\dag} \mathbf{V}_N^{\dag} \mathbf{T}\mathbf{V}_N\mathbf{w}_I},\\
\digamma&(n,p,x)\!=\!\left\{
\begin{array}{lll}
p^{n} \gamma\left(-n, p x\right), & n < 0,\\
\textbf{Ei}(-px), & n = 0,\\
\frac{p^{n} \textbf{Ei}(-px)}{n! (-1)^{n}}\!\!-\!\! \frac{e^{\!-\!px}}{x^{n}}\sum_{k=0}^{n-1}\frac{(-1)^kp^k x^k}{n(n\!-\!1)\dots (n\!-\!k)}, & n > 0,
\end{array}
\right.
\end{align}
$\gamma(\cdot, \cdot)$ is the incomplete gamma function, and $\textbf{Ei}(\cdot)$ is the exponential integral function.
\end{theorem}
\begin{IEEEproof}
Following \eqref{Pso_definition}, the exact secrecy outage probability of the CPA is given by
\begin{align}\label{Pso_CPA_definition}
P(R_s) = \Pr\left(\gamma_E \!>\! \frac{\gamma_B \!+\! 1}{2^{R_s}} \!-\! 1\right) = 1 \!-\! F_{\gamma_E}\left(\frac{\gamma_B \!+\! 1}{2^{R_s}} \!-\! 1\right),
\end{align}
where $F_{\gamma_E}(\cdot)$ is the cdf of $\gamma_E$. In order to derive the expression of $P(R_s)$, we next derive $F_{\gamma_E}(\cdot)$. We can rewrite $\gamma_E$ as $\gamma_E = \psi_n/\psi_d$, where
\begin{align}
\psi_n &= \alpha \mu_E |\mathbf{g}\mathbf{v}_{I}|^2, \\
\psi_d &= (1-\alpha) \mu_E
|\mathbf{g}
\mathbf{V}_{N}\mathbf{w}_I|^2 + 1.
\end{align}
When $\alpha = 1$, $\gamma_E = \mu_E |\mathbf{g}\mathbf{v}_{I}|^2$, and thus the cdf of $\gamma_E$ for $\alpha = 1$ is given by
\begin{align}
F_{\gamma_E}(z) = 1 - e^{-\frac{z}{\mu_E \mathbf{v}_I^{\dag} \mathbf{T}\mathbf{v}_I}}.
\end{align}
As such, $P(R_s)$ for $\alpha = 1$ is derived as given in \eqref{Pso_CPA_final}.
We note that $\mathbf{g}\mathbf{v}_I$ and $\mathbf{g}
\mathbf{V}_N\mathbf{w}_I$ are both complex Gaussian random variables (for given deterministic $\mathbf{v}_I$, $\mathbf{V}_{N}$, and $\mathbf{w}_I$). As such, for $0<\alpha<1$ the joint pdf of $\psi_n$ and $\psi_d$ is given by \cite{mallik2003on}
\begin{align}
f_{\psi_n, \psi_d}(x, y) = \frac{e^{-\frac{1}{2(1-\rho)}\left(\frac{x}{s_n^2}+\frac{y-1}{s_d^2}\right)}}{4 s_n^2 s_d^2 (1-\rho)}
I_0\left(\frac{\sqrt{\rho}}{1-\rho}\frac{\sqrt{x(y-1)}}{s_n s_d}\right).
\end{align}
Then, the pdf of $\gamma_E$ for $0<\alpha<1$ is derived as
\begin{align}\label{pdf_SINR_CPA}
f_{\gamma_E}(z)&= \int_0^{\infty} y f_{\psi_n, \psi_d}(zy, y) dy\notag\\
&\overset{e}{=}\frac{e^{\frac{1}{2(1-\rho)s_d^2}}}{4 s_n^2 s_d^2 (1-\rho)}\sum_{k=0}^{\infty}\frac{\left(\frac{\sqrt{\rho}}{1-\rho}\frac{\sqrt{z}}{s_n s_d}\right)^{2k}}{2^{2k}(k!)^2} \notag \\
& ~~~~\times \int_1^{\infty}y^{k + 1} (y-1)^k e^{-\frac{1}{2(1-\rho)}\left(\frac{z}{s_n^2}+\frac{1}{s_d^2}\right) y} dy\notag \\
&\overset{f}{=} \frac{e^{\frac{1}{2(1-\rho)s_d^2}}}{4 s_n^2 s_d^2 (1-\rho)}\sum_{k=0}^{\infty}\frac{\left(\frac{\sqrt{\rho}}{1-\rho}\frac{\sqrt{z}}{s_n s_d}\right)^{2k}}{2^{2k}(k!)^2} \sum_{i=0}^k \binom{k}{i} (-1)^{k-i}\notag \\
& ~~~~\times \int_1^{\infty}y^{k + i +1} e^{-\frac{1}{2(1-\rho)}\left(\frac{z}{s_n^2}+\frac{1}{s_d^2}\right) y} dy\notag \\
&\overset{g}{=} \frac{e^{\frac{1}{2(1-\rho)s_d^2}}}{4 s_n^2 s_d^2 (1-\rho)}\sum_{k=0}^{\infty}\frac{\left(\frac{\sqrt{\rho}}{1-\rho}\frac{\sqrt{z}}{s_n s_d}\right)^{2k}}{2^{2k}(k!)^2} \sum_{i=0}^k \frac{\binom{k}{i}}{ (\!-\!1)^{i\!-\!k}}\notag \\
& ~~~~\times \sum_{j = 0}^{k+i+1}\frac{ (k+i+1)! }{j!} \frac{e^{-\frac{1}{2(1-\rho)}\left(\frac{z}{s_n^2}+\frac{1}{s_d^2}\right)}}{ \left[\frac{1}{2(1\!-\!\rho)}\left(\frac{z}{s_n^2}\!+\!\frac{1}{s_d^2}\right)\right]^{k+i\!-\!j+2}},
\end{align}
where $\overset{e}{=}$ is achieved by noting $\psi_d \in [1, \infty)$, $\overset{f}{=}$ is obtained by applying \eqref{bio_expansion}, and $\overset{g}{=}$ is gained by utilizing the following identify
\begin{align}
\int_1^{\infty}x^n e^{- \tau x} dx = e^{- \tau}\sum_{j = 0}^n \frac{n!}{j!}\tau^{-n-j+1}.
\end{align}
Following \eqref{pdf_SINR_CPA}, the cdf of $\gamma_E$ for $0<\alpha<1$ is derived as
\begin{align}\label{cdf_SINR_CPA}
F_{\gamma_E}(z)&= \int_0^z f_{\gamma_E}(x) dx \notag \\
&= \frac{e^{\frac{1}{2(1-\rho)s_d^2}}}{4 s_n^2 s_d^2 (1-\rho)}\sum_{k=0}^{\infty}\frac{\rho^k  (1\!-\!\rho)^{-k}}{2^{k}(k!)^2 s_d^{2k}}
\sum_{i=0}^k \frac{\binom{k}{i}}{ (\!-\!1)^{i\!-\!k}}\notag \\
& ~~~~\times \sum_{j = 0}^{k+i+1}\frac{ (k+i+1)! }{j! \left[2(1\!-\!\rho)s_n^2\right]^{j\!-\!i\!-\!2}} \notag\\
&~~~~\times \int_0^z x^k \left(x \!+\! \frac{s_n^2}{s_d^2}\right)^{\!-\!(k+i\!-\!j+2)}e^{\!-\!\frac{1}{2(1\!-\!\rho)s_n^2}\left(x+\frac{s_n^2}{s_d^2}\right)} dx\notag\\
&\overset{h}{=} \frac{e^{\frac{1}{2(1-\rho)s_d^2}}}{4 s_n^2 s_d^2 (1-\rho)}\sum_{k=0}^{\infty}\frac{\rho^k  (1\!-\!\rho)^{-k}}{2^{k}(k!)^2 s_d^{2k}}
\sum_{i=0}^k \frac{\binom{k}{i}}{ (\!-\!1)^{i\!-\!k}}\notag \\
& ~~~~\times \sum_{j = 0}^{k+i+1}\frac{ (k+i+1)! }{j! \left[2(1\!-\!\rho)s_n^2\right]^{j\!-\!i\!-\!2}} \sum_{m = 0}^k\binom{k}{m}\left(\!-\!\frac{s_n^2}{s_d^2}\right)^{k\!-\!m}\notag\\
&~~~~\times \int_{\frac{s_n^2}{s_d^2}}^{z + \frac{s_n^2}{s_d^2}} y^{m\!-\!k\!-\!i+j\!-\!2}e^{\!-\!\frac{1}{2(1\!-\!\rho)s_n^2}y} dy\notag\\
&\overset{i}{=} \frac{e^{\frac{1}{2(1-\rho)s_d^2}}}{4 s_n^2 s_d^2 (1-\rho)}\sum_{k=0}^{\infty}\frac{\rho^k  (1\!-\!\rho)^{-k}}{2^{k}(k!)^2 s_d^{2k}}
\sum_{i=0}^k \frac{\binom{k}{i}}{ (\!-\!1)^{i\!-\!k}}\notag
\end{align}
\begin{align}
& ~~~~\times \sum_{j = 0}^{k+i+1}\frac{ (k+i+1)! }{j! \left[2(1\!-\!\rho)s_n^2\right]^{j\!-\!i\!-\!2}} \sum_{m = 0}^k\binom{k}{m}\left(\!-\!\frac{s_n^2}{s_d^2}\right)^{k\!-\!m}\notag\\
&~~~~\times \left[\digamma\left(k+i\!\!-\!\!j\!\!-\!\!m\!+\!1, \frac{1}{2(1\!-\!\rho)s_n^2}, z + \frac{s_n^2}{s_d^2}\right)\right.\notag \\
&~~~~~~~~~~ \left.- \digamma\left(k+i\!\!-\!\!j\!\!-\!\!m\!+\!1, \frac{1}{2(1\!-\!\rho)s_n^2}, \frac{s_n^2}{s_d^2}\right)\right],
\end{align}
where the incomplete gamma function $\gamma(\cdot, \cdot)$ is  given by
\begin{align}
\gamma(n, x) = \int_0^x e^{-t} t^{n-1} dt,
\end{align}
and  the exponential integral function $\textbf{Ei}(\cdot)$ is given by
\begin{align}
\textbf{Ei}(x) = -\int_{-x}^{\infty}\frac{e^{-t}}{t}dt = \int_{-\infty}^{x}\frac{e^{t}}{t}dt, ~~[x < 0].
\end{align}
We note that $\overset{h}{=}$ in \eqref{cdf_SINR_CPA} is achieved by exchanging random variable (i.e., $y = x + s_n^2/s_d^2$) and $\overset{i}{=}$ is gained with the aid of the following two identities
\begin{align}
&\int_0^{\tau} x^n e^{-px} dx = p^{-n-1}\gamma(n+1, p \tau), [n \geq 0], \\
&\int_{\tau}^{\infty} \frac{e^{-px} dx}{x^{n+1}} = \frac{p^{n} \textbf{Ei}(-p\tau)}{n! (-1)^{n+1}}\!\!-\!\! \frac{e^{\!-\!p\tau}}{\tau^{n}}\sum_{k=0}^{n-1}\frac{(-1)^kp^k \tau^k}{n(n\!-\!1)\dots (n\!-\!k)}, \notag \\
&~~~~~~~~~~~~~~~~~~~~~~~~~~~~~~~~~~~~~~~~~~~~~~~~~~[n > 0].
\end{align}
Then, the secrecy outage probability given in \eqref{Pso_CPA} is achieved by substituting \eqref{cdf_SINR_CPA} into \eqref{Pso_CPA_definition}.
\end{IEEEproof}

Comparing \eqref{Pso_CPA_final} and \eqref{asy_Pso_CPA}, we know that the expression for the exact secrecy outage probability is much more complicated than that for the asymptotic secrecy outage probability. Similar to \eqref{asy_Pso_CPA}, the infinite series involved in \eqref{Pso_CPA} can be approximated by its finite truncation accurately due to the convergence of  this infinite series.
As per \eqref{Pso_CPA_final}, we can draw the conclusion that when $\alpha = 1$ the exact secrecy outage probability averaged over all realizations of $\mathbf{h}$ increases as the correlation becomes more severe. This is due to the following two facts. Firstly, based on \eqref{SNR_Bob} the average value of $\gamma_B$ does not vary with the correlation matrix $\mathbf{T}$. Secondly, $\mathbf{v}_I^{\dag}\mathbf{T}\mathbf{v}_I^{\dag}$ increases on average as the correlation becomes more severe due to the fact that if $\mathbf{h}_s^{\dag}\mathbf{h}_s = \mathbf{I}_{N_t}$ we have $\mathbf{v}_I^{\dag}\mathbf{T}\mathbf{v}_I^{\dag} = \lambda_1$ and $\lambda_1$ increases as  the correlation becomes more severe. The detrimental effect of correlation can be intuitively understood as follows. As correlation increases, the eavesdropper's channel becomes more aligned with the main channel and thus the SNR at Eve improves in the absence of AN (since $\alpha = 1$). As we will confirmed later in Section V, this conclusion also holds with AN, i.e., when $\alpha$ is optimized.

\subsection{Optimization of the Power Allocation Parameter $\alpha$}

In order to guarantee the secrecy outage probability of a secure transmission is less than one, we have to guarantee $C_B > R_s$. As such, the lower bound on the power allocation parameter $\alpha$ is given by
\begin{align}
\alpha_l = \frac{2^{R_s}-1}{\mu_B \|\mathbf{h}\|^2}.
\end{align}

The optimal value of power allocation parameter $\alpha$ that minimizes the asymptotic secrecy outage probability given in \eqref{asy_Pso_CPA} can be obtained through
\begin{align}
\alpha_{\infty}^{\ast} =  \argmin_{\alpha_l < \alpha < 1} P_{\infty}(R_s),
\end{align}
and we denote $P_{\infty}^{\ast}(R_s)$ as the minimum asymptotic secrecy outage probability achieved by setting $\alpha = \alpha_{\infty}^{\ast}$ in the asymptotic secrecy outage probability $P_{\infty}(R_s)$.

The optimal value of power allocation parameter $\alpha$ that minimizes the exact secrecy outage probability given in \eqref{Pso_CPA_final} can be obtained through
\begin{align}
\alpha^{\ast} =  \argmin_{\alpha_l < \alpha \leq 1} P(R_s),
\end{align}
and we denote $P^{\ast}(R_s)$ as the minimum exact secrecy outage probability achieved by setting $\alpha = {\alpha}^{\ast}$ in the exact secrecy outage probability $P(R_s)$ .

\section{Numerical Results}\label{numerical}

In this section, we first provide numerical comparison among the OPA, CPA, and UPA, based on which we observe that our proposed CPA is nearly optimal in terms of minimizing the secrecy outage probability even for finite $N_t$. We also examine the impact of transmitter-side correlation on the AN-aided secure transmission. Finally, based on our numerical studies we draw useful insights on the effect of system parameters (e.g., $N_t$) on the AN-aided secure transmission with different power allocations.

\begin{figure}[!t]
    \begin{center}
   {\includegraphics[width=3.5in, height=2.9in]{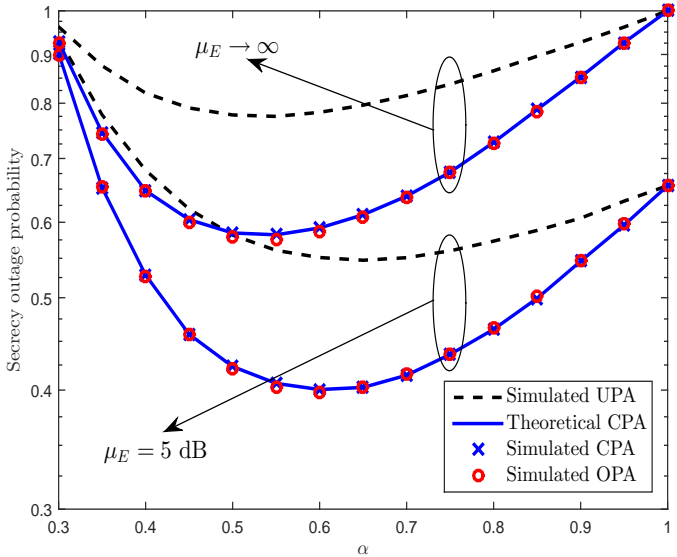}}
    \caption{Secrecy outage probabilities of the AN-aided secure transmission with OPA, CPA, and UPA versus different values of $\alpha$ for $N_t = 4$, $R_s = 2$, $\mu_B = 5$ dB, $\mathbf{\Lambda} = \text{diag}[2.8, 0.7, 0.3, 0.2]$, and $\mathbf{h}_s = [0.1104 \!-\! 0.6619i, \!-\!0.6677 + 1.2432i, 0.7588 + 0.9201i, 1.0196 + 0.4098i]$.}\label{fig:Pso_verify}
    \end{center}
\end{figure}

In Fig.~\ref{fig:Pso_verify} we plot the asymptotic (i.e., for $\mu_E \rightarrow \infty$) and exact (i.e., for $\mu_E = 5$ dB) secrecy outage probabilities of the AN-aided secure transmission with OPA, CPA, and UPA. In this figure we first observe that the theoretic secrecy outage probabilities achieved by the CPA precisely match its simulated secrecy outage probabilities. This confirms the correctness of the results presented in Theorem~\ref{theorem2} and Theorem~\ref{theorem3}. We note that in the calculation of our theoretic secrecy outage probabilities (i.e., $P_{\infty}(R_s)$ and $P(R_s)$) we truncate the infinite series $\sum_{k =0}^{\infty}$ at $k = 15$. As such, the exact match between the theoretic and simulated secrecy outage probabilities as shown in this figure also indicates that the theoretic $P_{\infty}(R_s)$ and $P(R_s)$ can be approximated accurately by closed-form expressions (i.e., the involved infinite series can be accurately approximated  by a finite series). Surprisingly, we also observe that our proposed CPA achieves almost identical secrecy outage probabilities as the OPA, which demonstrates that the proposed CPA is nearly optimal in terms of minimizing the secrecy outage probability under the adopted specific simulation settings (which are detailed in the caption of this figure\footnote{As we mentioned in Section II-B, $\mathbf{U}_T$ does not affect the performance of the AN-aided secure transmission. Thus, we only presented the adopted $\mathbf{\Lambda}$ for this figure.}). Noting $N_t = 4$ for Fig.~\ref{fig:Pso_verify}, we can conclude that our proposed CPA is nearly optimal even for a finite number of transmit antennas with moderate correlation. We note that as we have proved in Theorem~\ref{theorem1} the CPA is optimal in the large system regime with $N_t \rightarrow \infty$. In this figure, we also observe that the proposed CPA achieves much lower secrecy outage probabilities than the UPA, which shows one advantage of the CPA relative to the UPA.
Finally, we note that the optimal value of $\alpha$ that minimizes the secrecy outage probability for the CPA is smaller than that for the UPA (for $\mu_E \rightarrow \infty$ the gap is tiny, but when $\mu_E$ is finite, e.g.,  $\mu_E = 5$ dB, this gap is noticeable). This is mainly due to the fact that under the specific system settings the CPA utilizes AN more efficiently relative to the UPA (i.e., by using the same transmit power of AN, the CPA leads to more interference to Eve on average than the UPA). For $\mu_E \rightarrow \infty$, the optimal values of $\alpha$ are already low (considerable transmit power is allocated to AN) and thus the gap is relatively minor. For finite $\mu_E$, the optimal values of $\alpha$ are relatively high and thus the gap becomes noticeable.


\begin{figure}[!t]
    \begin{center}
   {\includegraphics[width=3.5in, height=2.9in]{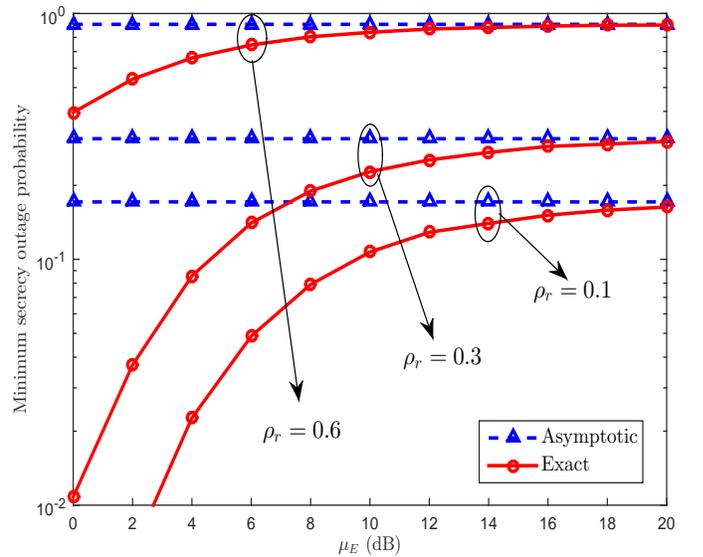}}
    \caption{Minimum secrecy outage probabilities achieved by the CPA versus different values of $\mu_E$ for $N_t = 3$, $L = 0.5$m (i.e., $\delta_{1,2} = \delta_{2,3}= 0.25$m and $\delta_{1,3} = 0.5$m), $R_s = 1$, $\mu_B = 10$ dB, and $\mathbf{h}_s = [0.0234 + 0.2351i,  \!-\!0.0788 \!-\! 0.4619i,  \!-\!0.5344 + 1.0582i]$.}\label{fig:Pso_SNR_E}
    \end{center}
\end{figure}

\begin{figure}[!t]
    \begin{center}
   {\includegraphics[width=3.5in, height=2.9in]{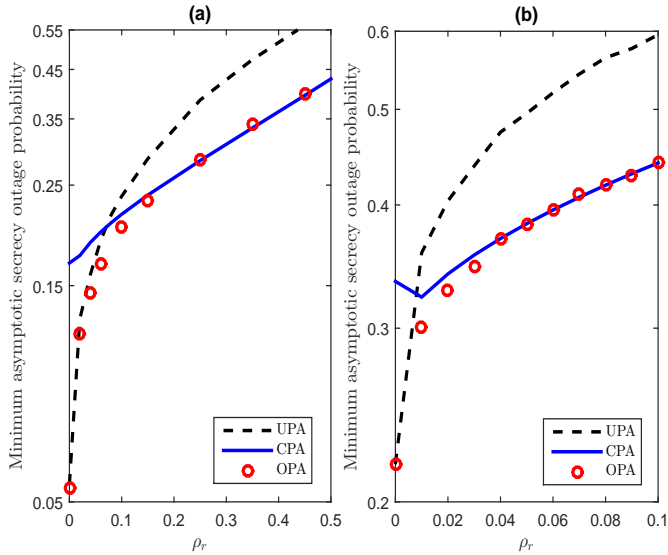}}
    \caption{Minimum asymptotic secrecy outage probabilities achieved by the OPA, CPA, and UPA versus different values of $\rho_r$ for $N_t = 3$, $L = 0.5$m, $R_s = 1$, $\mu_B = 10$ dB, and (a) $\mathbf{h}_s = [\!-\!0.1470 + 0.1876i,  \!-\!0.3905 + 1.0675i,  \!-\!0.5091 \!-\! 0.8150i]$, (b) $\mathbf{h}_s = [\!-\!0.0845 + 0.5064i,   0.1612 + 0.0330i,  \!-\!0.7529 + 0.0282i]$. }\label{fig:Pso_correlation_details}
    \end{center}
\end{figure}

\begin{figure}[!t]
    \begin{center}
   {\includegraphics[width=3.5in, height=2.9in]{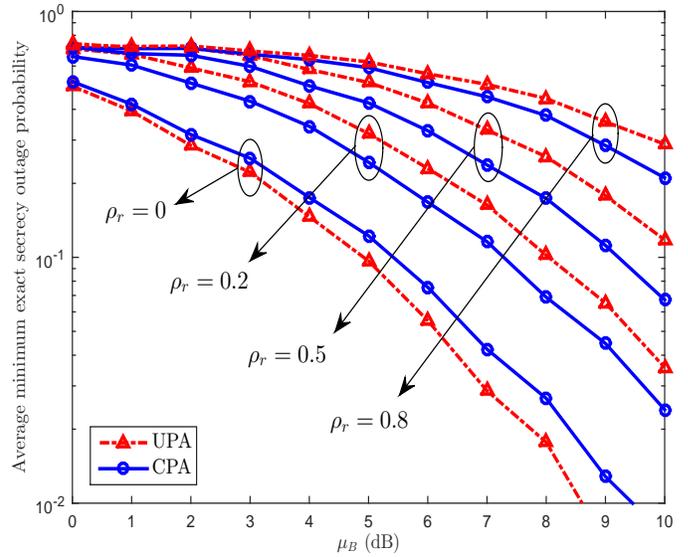}}
    \caption{Expected minimum exact secrecy outage probabilities achieved by the CPA and UPA versus different values of $\mu_B$ for $N_t = 3$, $L = 0.5$m, $R_s = 1$, $\mu_E = 5$ dB.}\label{fig:Ex_Pso_SNR_B}
    \end{center}
\end{figure}

In Fig.~\ref{fig:Pso_SNR_E} we plot the minimum asymptotic and exact secrecy outage probabilities of the CPA (i.e., $P_{\infty}^{\ast}(R_s)$ and $P^{\ast}(R_s)$) versus different values of $\mu_E$. Although our analysis is valid for an arbitrary correlation matrix, in this figure and the following figures we adopt an \emph{exponential correlation model}, in which the $(i,j)$-th entry of $\mathbf{T}$ is given by $t_{ij} = \rho_r^{\delta_{ij}}$, where $\rho_r \in [0,1]$ is the correlation parameter specified by system settings (e.g., signal frequency) and $\delta_{ij}$ is the distance between the $i$-th and $j$-th antennas at Alice. We note that a larger $\rho_r$ indicates a larger correlation for a fixed $\delta_{ij}$, where $\rho_r = 0$ serves as the uncorrelated case and $\rho_r = 1$ represents the fully correlated case. We also adopt the \emph{uniform linear array} as Alice's antenna configuration and the array length is denoted as $L$ in this figure and the following figures. In this figure we first observe that $P^{\ast}(R_s)$ approaches $P_{\infty}^{\ast}(R_s)$ as $\mu_E$ increases and they are almost equal when $\mu_E \geq 20$ dB. This demonstrates that $P_{\infty}^{\ast}(R_s)$ can be a good approximation of $P^{\ast}(R_s)$ even for a finite $\mu_E$ (e.g., 20 dB).
We also observe that for the adopted $\mathbf{h}_s$, both $P_{\infty}^{\ast}(R_s)$ and $P^{\ast}(R_s)$ increase as $\rho_r$ increases, which means that the secrecy performance of the CPA decreases as the correlation of transmit antennas becomes more severe. This is mainly caused by the fact that as the correlation increases the null space of the main channel shrinks. As this null space shrinks, the probability that the eavesdropper's channel lies in the null space decreases. As such, transmitting AN in the null space leads to less interference to Eve on average as the correlation increases.
We note that as the transmit antenna correlation increases the statistical CSI can better represent the instantaneous CSI \cite{wang2012statistical}. As we assume the statistical information on the eavesdropper's channel is known, the increase in the correlation offers more information on the instantaneous eavesdropper's channel. This seems to be in contradiction with the observation that both $P_{\infty}^{\ast}(R_s)$ and $P^{\ast}(R_s)$ increase as $\rho_r$ increases. The main reason is that the additional information offered by the increased correlation cannot counteract the decay in the null space of the main channel. We will further examine the impact of the transmitter-side correlation on the performance of the CPA in Fig.~\ref{fig:Pso_correlation_details} and Fig.~\ref{fig:Ex_Pso_SNR_B}, in which we will see that the increased correlation may decrease the secrecy outage probability of the CPA.

In Fig.~\ref{fig:Pso_correlation_details} we plot the minimum asymptotic secrecy outage probabilities achieved by the UPA, CPA, and OPA versus different values of $\rho_r$. In Fig.~\ref{fig:Pso_correlation_details}~(a) for the specific adopted $\mathbf{h}_s$ we observe that all the three minimum asymptotic secrecy outage probabilities increase as $\rho_r$ increases. This is due to the fact that as $\rho_r \rightarrow 1$ the null space of the main channel disappears and Alice cannot create interference to Eve while perfectly avoid the interference to Bob. In Fig.~\ref{fig:Pso_correlation_details}~(b) for the specific adopted $\mathbf{h}_s$ we observe that the minimum asymptotic secrecy outage probability achieved by the CPA first decreases and then increases as $\rho_r$ increases. This initial decrease can be explained by the fact that the initial increase in the correlation (i.e., $\rho_r$) offers useful information of the eavesdropper's channel while does not significantly affect the null space of the specific $\mathbf{h}_s$, which leads to the reduction in the secrecy outage probability.
The following increase in the minimum asymptotic secrecy outage probability achieved by the CPA can be explained by the fact that the offered information on the eavesdropper's channel by the increase in $\rho_r$ cannot counteract the decay of the null space caused by the increase in $\rho_r$.  We conducted hundreds of simulations for different realizations of $\mathbf{h}_s$ and all the results are similar to either Fig.~\ref{fig:Pso_correlation_details}~(a) or Fig.~\ref{fig:Pso_correlation_details}~(b). In both Fig.~\ref{fig:Pso_correlation_details}~(a) and Fig.~\ref{fig:Pso_correlation_details}~(b) we also observe that the minimum asymptotic secrecy outage probability achieved by the CPA is lower than that achieved by the UPA when $\rho_r$ is larger than some specific value, which only corresponds to a small correlation.

As we mentioned in Section II-C, we have focused on the secrecy outage probability for a given main channel realization in order to study the AN power allocation design based on a given main channel realization. Now, we also present the secrecy outage probability averaged over all main channel realizations.
In Fig.~\ref{fig:Ex_Pso_SNR_B} we plot the average minimum exact secrecy outage probabilities achieved by the CPA and UPA, which are obtained through averaging the minimum exact secrecy outage probabilities over $\mathbf{h}$ obtained by the CPA and UPA, respectively. As expected, in this figure we first observe that the average minimum exact secrecy outage probability achieved by the CPA is higher than that achieved by the UPA for the uncorrelated case (i.e., $\rho_r = 0$). This is due to the fact that for $\rho_r = 0$ the variance of the interference to Eve for the CPA is larger than that for the UPA while the mean of the interference to Eve for these two power allocations is the same. We also observe that the average minimum exact secrecy outage probability achieved by the CPA is lower than that achieved by the UPA for the most of the correlated cases (e.g., for $\rho_r \geq 0.2$). This can be explained by the fact that on average the CPA leads to a much larger average interference to Eve relative to the UPA. The aforementioned two observations again demonstrate that our proposed CPA can outperform the UPA in correlated fading channels. Finally, as expected we observe that both these two average minimum exact secrecy outage probabilities decrease as $\mu_B$ increases, which caused by the increasing quality of the main channel.

\begin{figure}[!t]
    \begin{center}
   {\includegraphics[width=3.5in, height=2.9in]{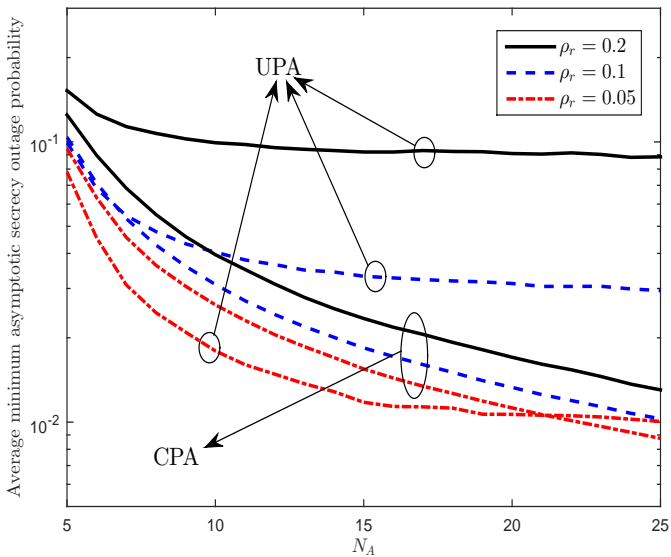}}
    \caption{Average minimum asymptotic secrecy outage probabilities achieved by the CPA and UPA versus different values of $N_t$ for $L = 0.5$m, $R_s = 1$, $\mu_B = 10$ dB, and $\mu_E = 5$ dB.}\label{fig:Ex_Pso_N_t}
    \end{center}
\end{figure}


Continuing with Fig.~\ref{fig:Ex_Pso_SNR_B}, we plot the average minimum asymptotic secrecy outage probabilities achieved by the CPA and UPA versus different values of $N_t$ for fixed array length $L$ in Fig.~\ref{fig:Ex_Pso_N_t} (Alice is equipped with uniform linear array with $N_t$ elements and the array length is $L$). In this figure, we first observe that the average minimum asymptotic secrecy outage probability achieved by the UPA first decreases and then keeps nearly constant as $N_t$ increases. This indicates that $N_t$ suffers from a saturation point in improving the secrecy performance of the UPA, which can be explained by the fact that as $N_t$ increases for a fixed $L$ the correlation among these transmit antennas becomes stronger. On the contrary, we observe that the average minimum asymptotic secrecy outage probability achieved by the CPA continuously decreases as $N_t$ increases without such a saturation point. This demonstrates another advantage of our proposed CPA relative to the UPA, which is that when $N_t$ is large the CPA outperforms the UPA even in the wiretap channel with very low transmitter-side correlation. This advantage is confirmed by the observation that the average minimum asymptotic secrecy outage probability achieved by the CPA becomes lower than that achieved by the UPA for $\rho_r = 0.05$  when $N_t$ is larger than 21. Also, this observation can be explained by our Theorem~\ref{theorem1}, in which we have proved that the CPA is optimal in the large system-regime.
Finally, we observe that the gap between these two average minimum asymptotic secrecy outage probabilities increases as $N_t$ increases.

\section{Conclusion}\label{conclusion}

In this work, we devised the CPA for AN in the wiretap channel with transmitter-side correlation and theoretically proved its optimality in terms of achieving the minimum secrecy outage probability in the large system regime with $N_t \rightarrow \infty$. Our analysis showed that the proposed CPA maximizes the average interference to Eve for arbitrary $N_t$. The asymptotic and exact secrecy outage probabilities of the CPA for arbitrary $N_t$ were derived in easy-to-evaluate expressions, based on which we optimized the power allocation between the information signal and AN in the AN-aided secure transmission with CPA.
The conducted numerical results demonstrated that the CPA is nearly optimal and significantly outperforms the UPA  even for finite $N_t$ in the wiretap channel with transmitter-side correlation. Our study also revealed that the secrecy outage probability achieved by the UPA suffers from a saturation point as $N_t$ increases while the secrecy outage probability obtained by our proposed CPA does not.

%

\vspace{-1cm}
\begin{IEEEbiography}[{\includegraphics[width=1in,height=1.25in,clip,keepaspectratio]{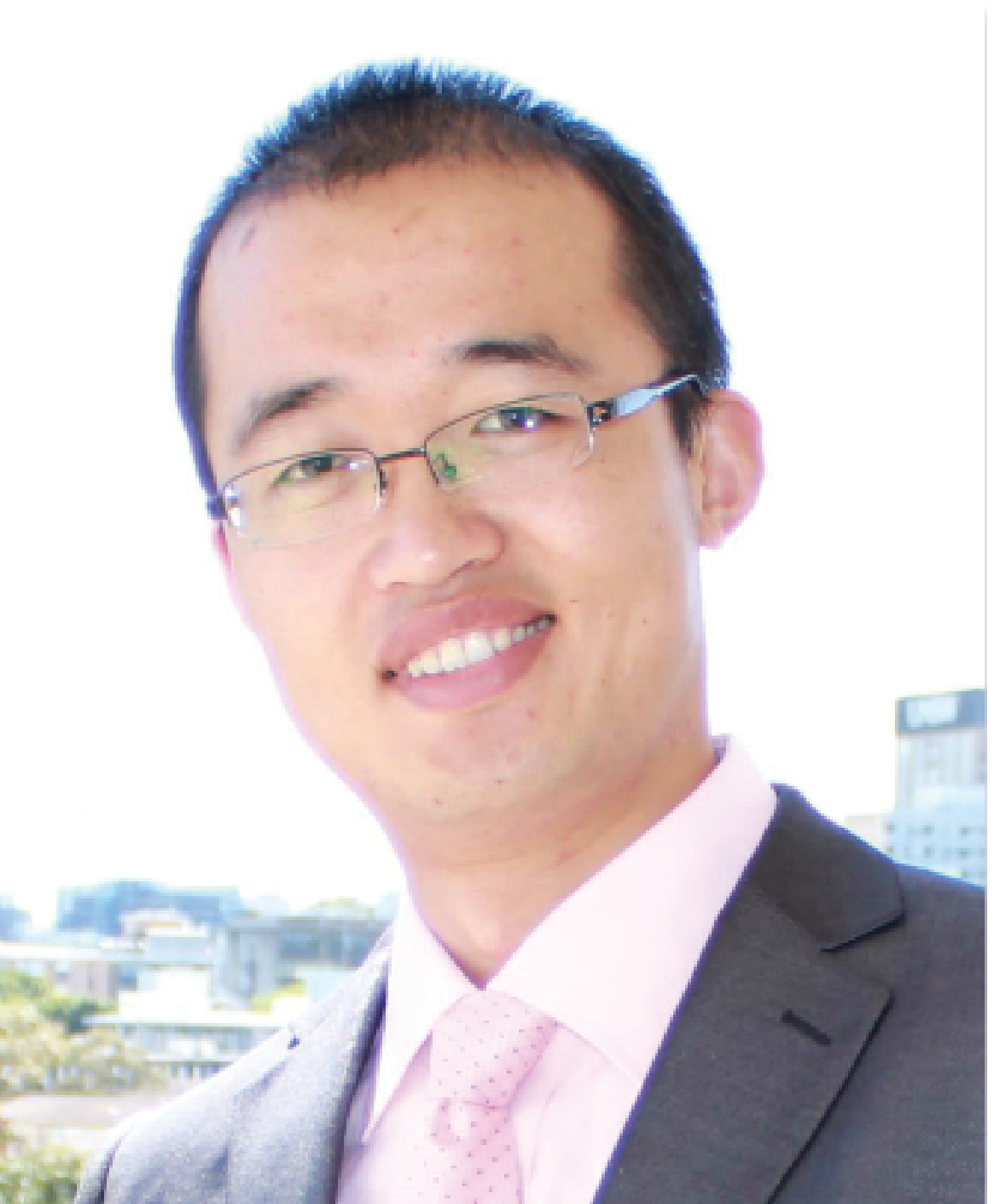}}]
{Shihao Yan} (S'11-M'15) received the Ph.D degree in Electrical Engineering from The University of New South Wales, Sydney, Australia, in 2015. He received the B.S. in Communication Engineering and the M.S. in Communication and Information Systems from Shandong University, Jinan, China, in 2009 and 2012, respectively. He was a visiting Ph.D student at The University of South Australia, Adelaide, Australia, in
2014. He is currently a Postdoctoral Research Fellow in the Research School of Engineering, The Australia National University, Canberra, Australia. His current research interests are in the areas of wireless communications and statistical signal processing, including physical layer security, location verification, and localization algorithms.
\end{IEEEbiography}

\vspace{-0.5cm}
\begin{IEEEbiography}[{\includegraphics[width=1in,height=1.25in,clip,keepaspectratio]{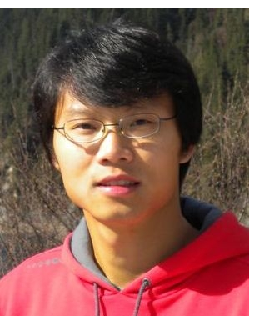}}]
{Xiangyun Zhou} (M'11)
received the Ph.D. degree in telecommunications engineering from the Australian National University (ANU) in 2010. From 2010 to 2011, he worked as a postdoctoral fellow at UNIK - University Graduate Center, University of Oslo, Norway. He returned to ANU in 2011 and currently works as a Senior Lecturer. His research interests are in the fields of communication theory and wireless networks. Dr. Zhou currently serves on the editorial board of IEEE Transactions on Wireless Communications and IEEE Communication Letters. He also served as a guest editor for IEEE Communication Magazine's feature topic on wireless physical layer security in 2015 and EURASIP Journal on Wireless Communications and Networking's special issue on energy harvesting wireless communications in 2014. He has also served as symposium/track/workshop co-chairs for major IEEE conferences. He was the chair of the ACT Chapter of the IEEE Communications Society and Signal Processing Society from 2013 to 2014. He is a recipient of the Best Paper Award at ICC'11.
\end{IEEEbiography}

\begin{IEEEbiography}[{\includegraphics[width=1in,height=1.25in,clip,keepaspectratio]{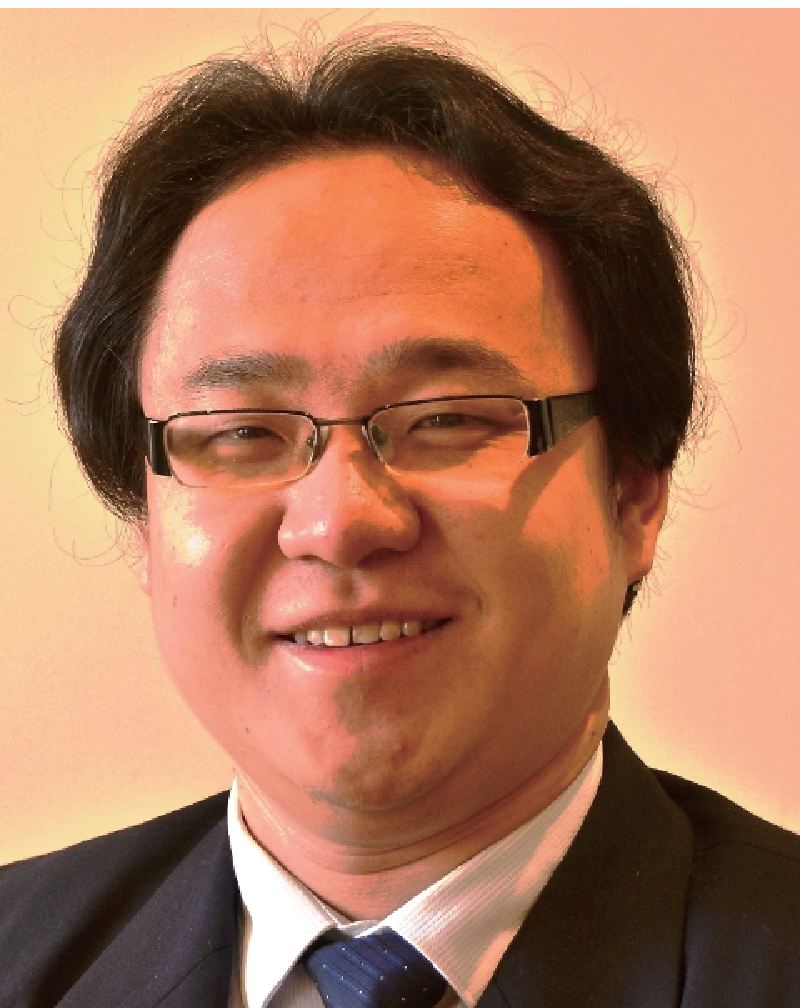}}]
{Nan Yang} (S'09--M'11) received the Ph.D. degree in electronic engineering from the Beijing Institute of Technology in 2011. Currently he is a Future Engineering Research Leadership Fellow and Lecturer in the Research School of Engineering at the Australian National University. Prior to this, he was a Postdoctoral Research Fellow at the University of New South Wales (UNSW) from 2012 to 2014 and a Postdoctoral Research Fellow at the Commonwealth Scientific and Industrial Research Organization from 2010 to 2012. He received the Exemplary Reviewer Award of the IEEE Transactions on Communications and the Top Reviewer Award from the IEEE Transactions on Vehicular Technology in 2015, the IEEE ComSoc Asia-Pacific Outstanding Young Researcher Award and the Exemplary Reviewer Award of the IEEE Wireless Communications Letters in 2014, the Exemplary Reviewer Award of the IEEE Communications Letters in 2013 and 2012, and the Best Paper Award at the IEEE 77th Vehicular Technology Conference in 2013. He is currently serving on the Editorial Board of the IEEE Transactions on Vehicular Technology and the Transactions on Emerging Telecommunications Technologies. His research interests include heterogeneous networks, massive multi-antenna systems, millimeter wave communications, cyber-physical security, and molecular communications.
\end{IEEEbiography}
\vspace{-7.5cm}
\begin{IEEEbiography}[{\includegraphics[width=1in,height=1.25in,clip,keepaspectratio]{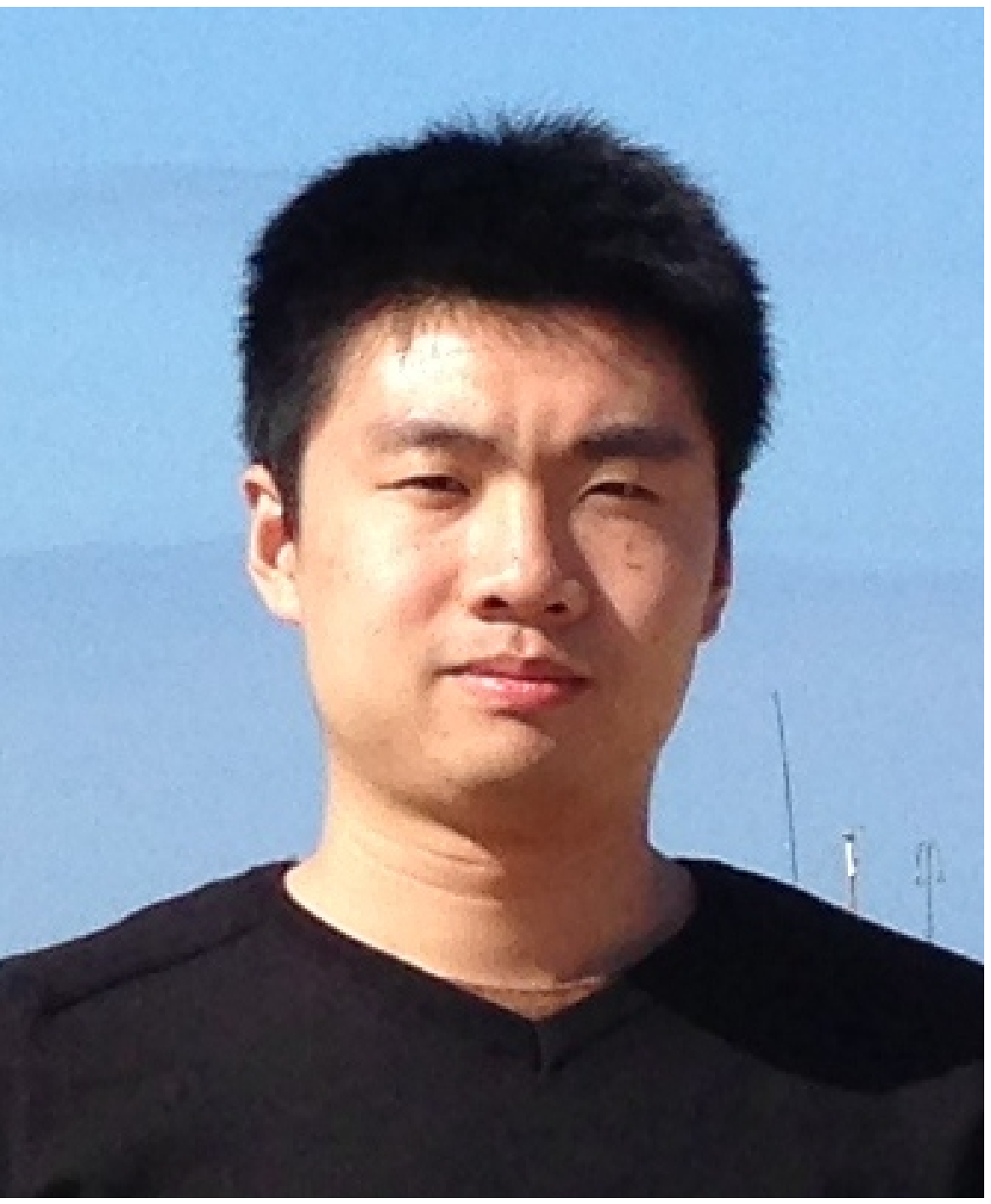}}]
{Biao He} (M'16)
received the B.E. (hons.) degree from the Australian National University (ANU) and Beijing Institute of Technology in 2012. He received the Ph.D. degree from the ANU in 2016. He is currently a Post-Doctoral Fellow at the Department of Electronic and Computer Engineering, The Hong Kong University of Science and Technology. His research interests include physical layer security and wireless communications.
\end{IEEEbiography}
\vspace{-7.5cm}
\begin{IEEEbiography}[{\includegraphics[width=1in,height=1.25in,keepaspectratio]{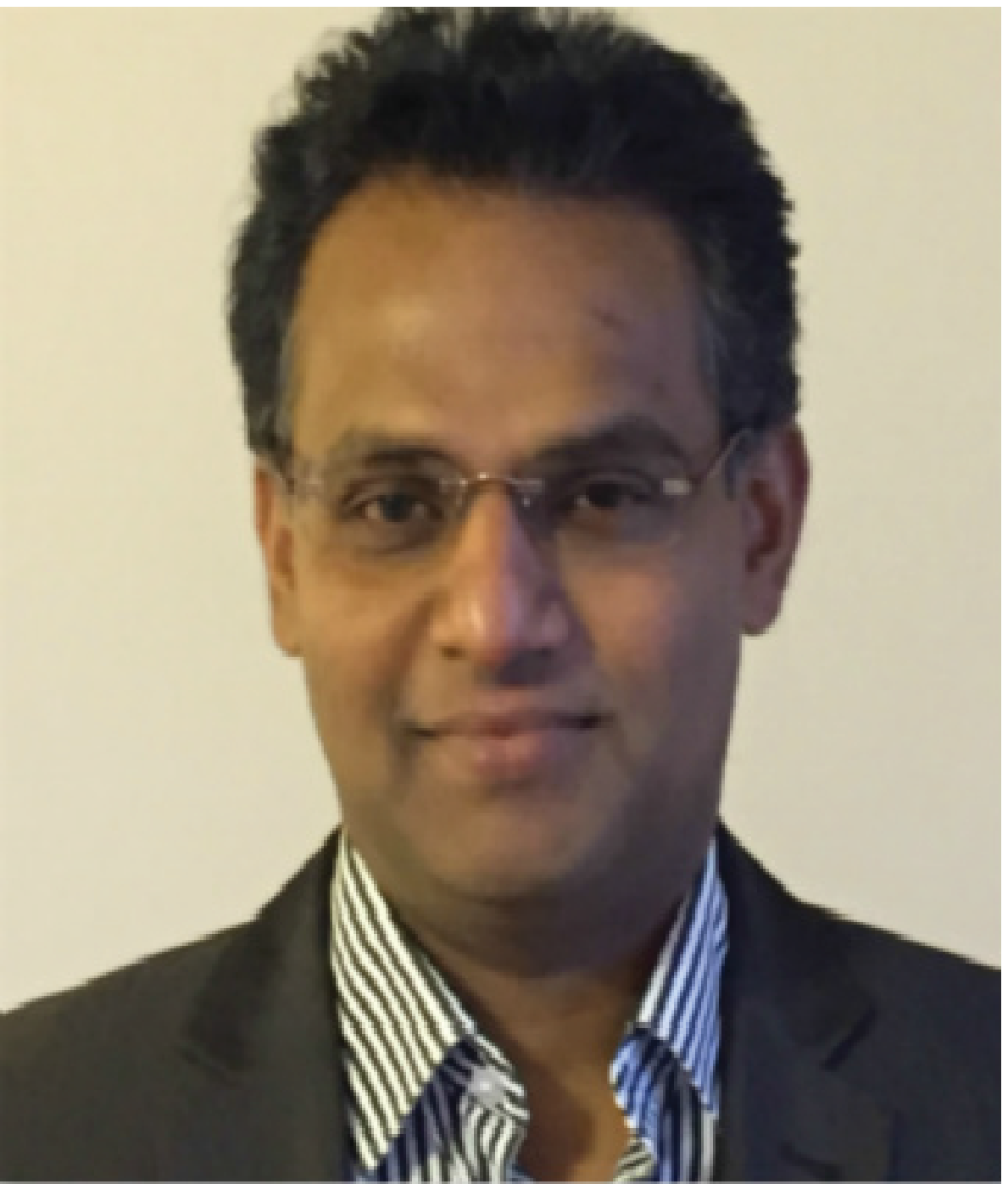}}]
{Thushara Abhayapala} (M'00--SM'08)  received the B.E. degree (with Honors) in Engineering in 1994 and the Ph.D. degree in Telecommunications Engineering in 1999, both from the Australian National University (ANU), Canberra. Currently he is the Deputy Dean of the College of Engineering \& Computer Science, ANU. He was the the Director of the Research School of Engineering  at ANU from January 2010 to October 2014 and the Leader of the Wireless Signal Processing (WSP) Program at the National ICT Australia (NICTA) from November 2005 to June 2007. His research interests are in the areas of spatial audio and acoustic signal processing, multi-channel signal processing and spatial aspects of wireless communications. He has supervised over 30 PhD students and coauthored over 200 peer reviewed papers. Professor Abhayapala is an Associate Editor  of IEEE/ACM Transactions on Audio, Speech, and Language Processing. He is a Member of the Audio and Acoustic Signal Processing Technical Committee (2011­-2016) of the IEEE Signal Processing Society. He is a Fellow of the Engineers Australia (IEAust).
\end{IEEEbiography}

\end{document}